\documentclass[twocolumn]{aastex7}
\usepackage{graphicx}
\pdfoutput=1 
\usepackage{amssymb}
\usepackage{gensymb}
\usepackage[caption=false]{subfig}
\usepackage{booktabs}
\usepackage{xcolor}
\usepackage{multirow}
\usepackage{natbib}
\usepackage{xspace}
\usepackage{comment}

\defcitealias{Moran2023}{Moran \& Stevenson et al. 2023}

\newcommand{\fleck}{\href{https://github.com/bmorris3/fleck}{\textcolor{cyan}{\tt fleck}}\xspace}

\begin{document}

\title{The HST/WFC3 Transmission Spectrum of AU Mic b Part I: An Atmosphere Obscured by Contamination and Systematics}

\author[0000-0002-8961-0352]{William~C.~Waalkes}
\affiliation{Department of Physics and Astronomy, Dartmouth College, Hanover NH 03755, USA}
\affiliation{Max Planck Institute for Solar System Research, Justus-von-Liebig-Weg 3, 37077 G{\"o}ttingen, Germany}
\email{waalkes@mps.mpg.de}

\author[0000-0002-8518-9601]{Peter Gao}
\affiliation{Earth and Planets Laboratory, Carnegie Institution for Science, Washington DC, USA}
\email{pgao@carnegiescience.edu}

\author[0000-0003-4150-841X]{Elisabeth Newton}
\affiliation{Department of Physics and Astronomy, Dartmouth College, Hanover NH 03755, USA}
\email{Elisabeth.R.Newton@dartmouth.edu}

\author[0000-0003-2528-3409]{Brett M.~Morris}
\affiliation{Space Telescope Science Institute, Baltimore, Maryland, USA}
\email{bmmorris@stsci.edu}

\author[0000-0002-3321-4924]{Zachory~K.~Berta-Thompson}
\affiliation{Department of Astrophysical \& Planetary Sciences, University of Colorado Boulder, 2000 Colorado Ave, Boulder, CO 80309, USA}
\email{zach.bertathompson@colorado.edu}

\author[0000-0003-4328-3867]{Hannah R. Wakeford}
\affiliation{School of Physics, University of Bristol, Bristol, UK}
\email{hannah.wakeford@bristol.ac.uk}

\author[0000-0001-8703-7751]{Lili Alderson}
\affiliation{Department of Astronomy, Cornell University, 122 Sciences Drive, Ithaca, NY 14853, USA}
\email{lili.alderson@cornell.edu}

\author[0000-0003-3654-1602]{Andrew W. Mann}
\affiliation{Department of Physics and Astronomy, The University of North Carolina at Chapel Hill, Chapel Hill, NC 27599-3255, USA}
\email{awmann@unc.edu}

\author[0000-0002-8864-1667]{Peter Plavchan}
\affiliation{Department of Physics and Astronomy, George Mason University, Fairfax, VA 22030, USA}
\email{pplavcha@gmu.edu}

\author[0000-0001-8014-0270]{Patrick J. Lowrance}
\affiliation{IPAC, California Institute of Technology, Pasadena CA 91125}
\email{lowrance@ipac.caltech.edu}

\author[0000-0003-1240-6844]{Natasha E. Batalha}
\affiliation{NASA Ames Research Center, Moffett Field, CA 94035, USA}
\email{natasha.e.batalha@nasa.gov}

\author{Eric D. Lopez}
\affiliation{NASA Goddard Space Flight Center, Greenbelt, MD 20771}
\email{eric.d.lopez@nasa.gov}

\author[0000-0003-3444-5908]{Roxana Lupu}
\affiliation{Eureka Scientific, Inc, Oakland, CA 94602}
\email{roxifera@gmail.com}

\begin{abstract}

Young sub-Neptune progenitors around M dwarfs offer an excellent opportunity to probe the formation of their abundant, older cousins. At $\sim$20 Myr and only 9.7 pc away, AU Mic b is an ideal candidate for this effort, with its density and observations of escaping hydrogen pointing to a significant primordial atmosphere.
Here we present the 0.8-1.6 $\micron$ transmission spectrum of AU Mic b observed with the Wide Field Camera 3 on the Hubble Space Telescope (HST). We find that HST experienced unstable scanning during its visits, resulting in a variable PSF that dramatically affects the orbit-to-orbit baseline of the observations.
While we were able to somewhat mitigate this problem through spectral binning, the effects cannot be completely eliminated, limiting the precision of our results. Our data is further impacted by the intense magnetic activity of AU Mic, which introduced significant rotational variability along with spot crossings and the transit light source (TLS) effect into the light curves and spectrum, respectively.
Through decomposition of the out-of-transit stellar SED, we are able to constrain AU Mic's photospheric and spot temperatures to 3891$\pm$37 and 3020$\pm$69 K, respectively, with a spot filling factor of $0.33\pm0.05$.
Using Bayesian atmospheric retrievals, we show that the spectrum is dominated by the TLS effect with weak atmospheric constraints, with the data preferring a relatively small scale height of $<$185 km to 3$\sigma$.
Extrapolation of our retrieved spectra shows that the TLS effect dominates over atmospheric features at optical and infrared wavelengths.

\end{abstract}

\section{Introduction} \label{intro}

Sub-Neptunes are an abundant and enigmatic population with no analogue in the Solar System, which have motivated significant effort into understanding their origins \citep{Bean2021}.
In particular, recent work has shown that some sub-Neptunes orbiting M dwarfs may be fundamentally different from those around FGK stars in terms of their bulk composition and evolution \citep{Luque2022,Ho2024}.
Sub-Neptunes around M dwarfs are also ideal targets for atmospheric characterization thanks to the larger star-to-planet radius ratios compared to FGK systems \citep{Kempton2023,Madhusudhan2023,Benneke2019}.
However, inferring formation processes from atmospheric observations of mature M dwarf sub-Neptunes could be hindered by extensive atmospheric loss and atmosphere-interior interactions over Gyr timescales \citep{Cherubim2025,Nixon2025,Werlen2025a,Werlen2025b,Valatsou2026, Steinmeyer2026}. 

\begin{table*}[hbt]
    \centering
    \begin{tabular}{cccccccc}
    \hline
    \textbf{Observation Date} & \textbf{Label} & \textbf{Grism} & \textbf{Bandpass} & \textbf{$\Delta\lambda$} & \textbf{Exposure Time} & \textbf{$\rm N_{times}$} & \textbf{$\rm N_{wavelengths}$}\\
    \hline
    2021 August 30 & F21 & G141 & 1.14-1.64 $\mu$m & 0.00464 $\mu$m & 4.9784 s & 162 & 110 \\
    2022 April 14 & S22 & G102 & 0.79-1.13 $\mu$m & 0.00246 $\mu$m & 9.6763 s & 160 & 138 \\
    \hline
    \end{tabular}
    \caption{Details on the two HST visits in this program. $\rm N_{times}$ is the per-scan and per-wavelength number of exposures. The start and end of either bandpass is based on the extracted wavelengths in the data processing step but does not represent the precise range of wavelengths after binning the lightcurves, at which point some end points were thrown out.}
    \label{tab:1-observations}
\end{table*}

Young ($<$1 Gyr) exoplanets represent a rarer but more pristine population for studying planet formation and early evolutionary processes.
Specifically, planets with ages $<$100 Myr may be at their least evolved form before significant atmospheric loss via photoevaporation \citep{Owen2012,Owen2013,Owen2017}, boil-off \citep{Owen2016}, and core-powered mass loss \citep{ Ginzburg2016, Gupta2019, Owen2019}.
Observing sub-Neptune progenitors around young M dwarfs could therefore offer an important pathway towards understanding the properties of their older cousins.
However, young stars, and young M dwarfs in particular, are extremely active and variable, impacting both the transit light curve and the transmission spectrum of their planets, the latter through the Transit Light Source (TLS) effect \citep{Zhang2018,Rackham2018,Wakeford2019,Rackham2019, Lim2023,RackhamdeWit2024,SeagerShapiro2024}.
For low-mass stars, their cool photospheres and spots can imprint steep optical slopes and water vapor spectral features onto the planetary transmission spectrum, leading to false positive detections of haze and water vapor \citep{Iyer2020, Barclay2021,Moran2023}.
Correcting for the TLS effect is complex due to observational degeneracies between spot sizes, morphologies, temperatures, and locations \citep[see e.g.][ for comprehensive overviews of spots]{Solanki2003,Berdyugina2005}.
As such, while observations of young M dwarf sub-Neptune precursors can be vital for understanding the population of mature sub-Neptunes orbiting cool stars, interpretation of their transmission spectra requires caution.

AU Mic is a nearby (9.7 pc), young ($\sim$20 Myr), bright (V=8.627) M0Ve pre-main sequence star and a member of the $\beta$ Pictoris moving group \citep{Mamajek2014}.
It is known to host an extensive debris disk \citep{Kalas2004} and at least two confirmed transiting planets \citep{Plavchan2020,Martioli2021}, with two additional unconfirmed candidates \citep{Wittrock2023,Donati2025}.
AU Mic b is the innermost of the confirmed planets, with a radius measured between 3.55$\pm$0.13 R$_{\Earth}$ \citep{Szabo2022} and 4.79$\pm$0.29 R$_{\Earth}$ \citep{Mallorquin2024}.
This planet's mass is uncertain, with estimates ranging ranging from $7\pm2$ M$_{\Earth}$ measured by radial velocity \citep{Donati2025} to $20\pm2$ M$_{\Earth}$ measured from transit-timing variation \citep{Cale2021}.
Given known difficulties with mass measurement for planets orbiting PMS stars \citep[e.g., ][]{Tran2021,Blunt2023}, we are inclined to place more trust in the TTV analysis, in line with the findings of \citet{Livingston2026} which used TTVs to rule out the RV-derived mass for a similar system, V1298 Tau b \citep{Mascareno2021}.
The transmission spectrum itself can also be used to measure planetary masses \citep{deWit2013}, but requires a strong detection of molecular features from which the atmospheric scale height can be constrained.
This has been done for a handful of systems similar to AU Mic like V1298 Tau b and c \citep{Barat2024a,Barat2024b,Barat2025} and HIP 67522b \citep{Thao2024}.

\begin{figure*}[ht!]
    \subfloat{\includegraphics[width=\textwidth]{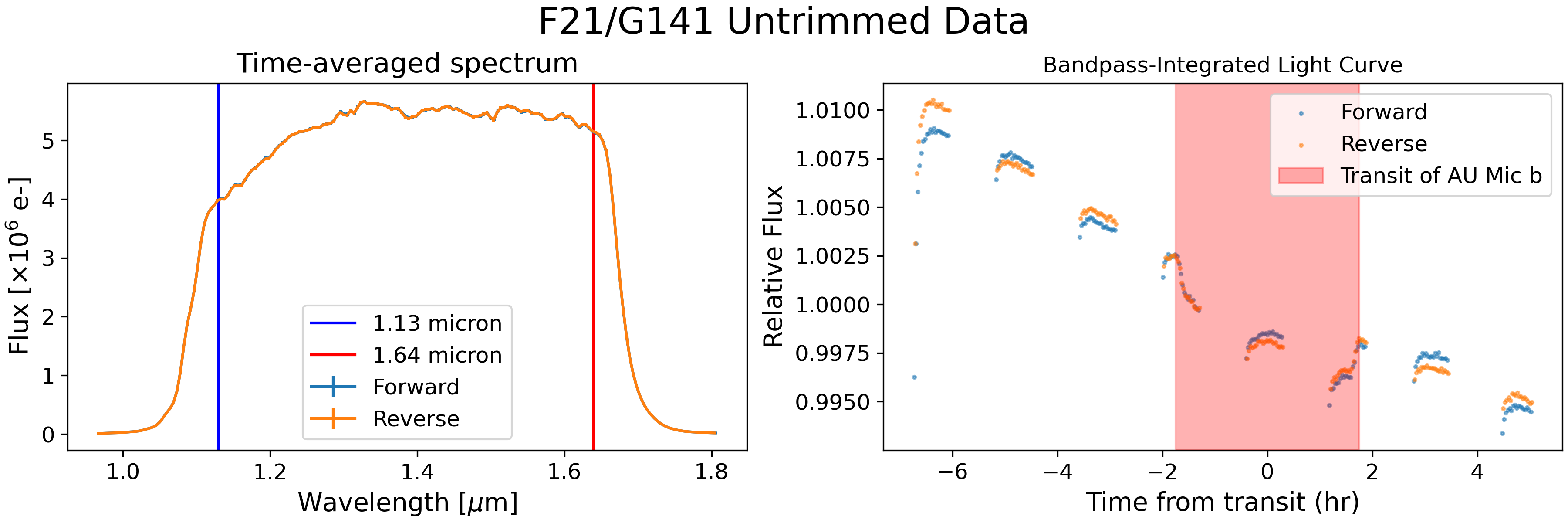}} \\
    \subfloat{\includegraphics[width=\textwidth]{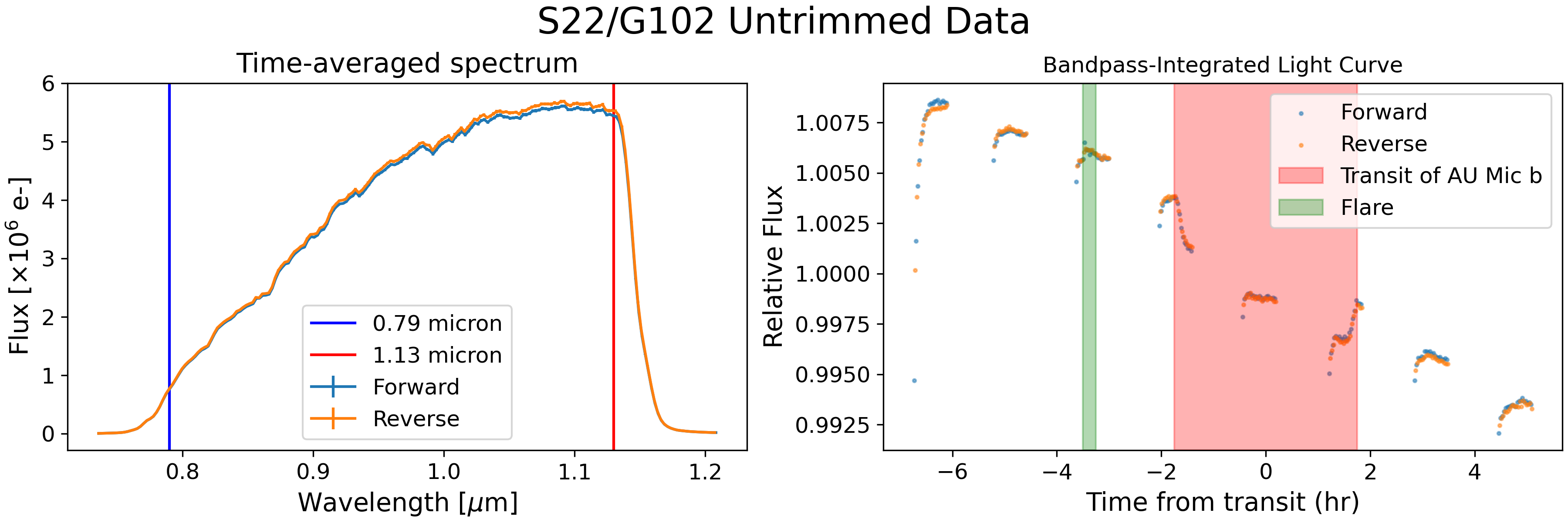}}
\caption{Median spectra and white light curves for the first (top row) and second visit (bottom row) with trim regions indicated on the left panel. The transit light curves show the typical WFC3 systematic ramp effect, with the first orbit in each visit experiencing the most dramatic effect. At least one flare is clear in the data, highlighted in green in the 2nd orbit of S22/G102. The flux offset between scan directions is more noticeable in the G141 observations.}
\label{fig:untrimmed_data}
\end{figure*}

With a semi-major axis of only 0.065 AU, AU Mic b experiences intense stellar winds and XUV irradiation that could significantly impact the survival of its atmosphere over time \citep{Carolan2020,Modi2023,Louca2023}.
However, the detection of variable neutral hydrogen escaping the planet by \citet{Rockcliffe2023} suggests that the atmosphere is still likely dominated by a primordial composition.
Therefore, AU Mic b is an ideal target for probing the near-initial conditions of sub-Neptune evolution around M dwarfs.

This planet is one of a very small number of nearby young sub-Neptunes which have been studied in depth, following HST observations of the $\sim$10 Myr sub-Neptune K2-33b \citep{Thao2023} and both HST and JWST observations of the $\sim$20-30 Myr planets V1298 Tau b and c \citep{Barat2024a, Barat2024b, Barat2025}.
K2-33b, which orbits a nearby, heavily spotted M dwarf, shows rising transit depths at bluer wavelengths.
\citet{Thao2023} found that the spot coverage fraction required to explain the blue-optical transit depths was $\sim60\%$, much higher than what was previously measured by spectral decomposition ($\leq20\%$; \citet{Thao2023}).
They suggested that the signal is instead due to Rayleigh scattering of photochemical hazes.
Both V1298 Tau b and c were found to have solar or less atmospheric metallicity, which implies significant atmospheric evolution probably awaits them over the next $\sim100$Myr of the system's age \citep{Barat2024b}.

\begin{figure*}[ht!]
\centering
    \includegraphics[width=0.97\textwidth]{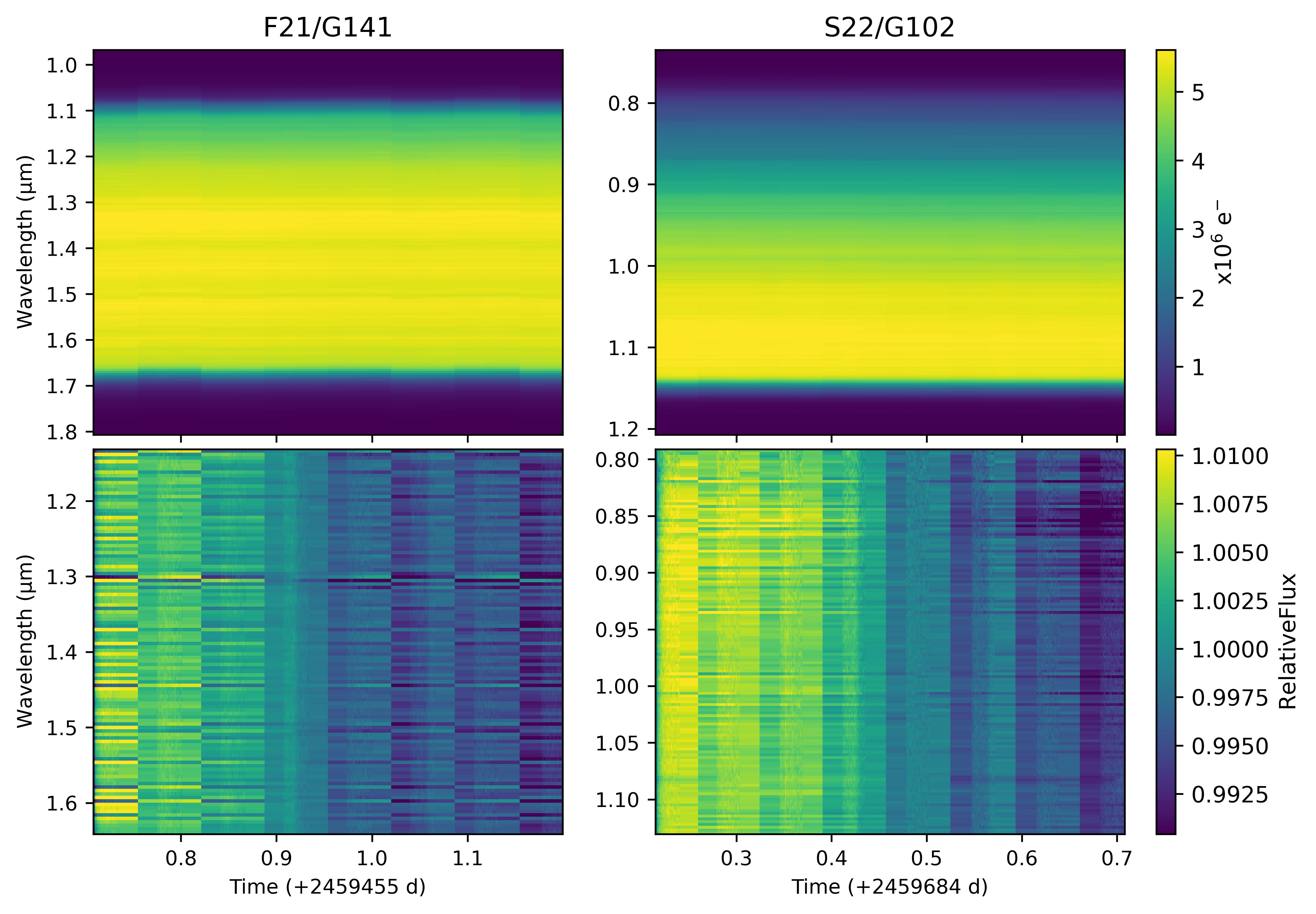}
\caption{{\bf Top row}: spectroscopic light curves for visit F21/G141 (left column) and S22/G102 (right column) as they are returned from the PACMAN pipeline, with the temporal gaps between orbits removed. 
{\bf Bottom row}: after trimming the wavelength edges and mean-normalizing each light curve, the effect of poor tracking stands as horizontal stripes of alternating bright and dark intensity, due to spectral lines wobbling in and out of the wavelength bin.
Observations from visit F21/G141 noticeably show a worse effect than S22/G102, but the same issue is present in both visits.}   
\label{fig:2Dspec}
\end{figure*}

Here we present the 0.8-1.6 $\mu$m transmission spectrum of AU Mic b observed by the \textit{Hubble Space Telescope} Wide Field Camera 3 G102 and G141 grisms, which captures the impact of both atmospheric absorption by diagnostic molecules like H$_2$O and CH$_4$ and the TLS effect.
We also investigate the out-of-transit stellar SED (OOT SED) and the rotation signal in the light curve to constrain the temperature and covering fraction of spots on AU Mic, which we then apply to atmospheric retrievals of AU Mic b's spectrum.
This work is the first of a two-part analysis.
In Paper I, presented here, we interpret the spot contaminated spectrum.
In Paper II, we construct an empirical model of the stellar photosphere to decontaminate the transmission spectrum.

We outline the observations and data reduction process in Section \ref{observations_and_data} and describe our methods for analyzing the light curves and extracting and interpreting the transmission spectrum in Section \ref{methods}.
We show our results for the analysis of the OOT SED and the atmospheric retrievals in Section \ref{results}, while in Section \ref{discussion} we discuss the caveats of our analysis and the implications for the nature of AU Mic b.
We present our conclusions in Section \ref{conclusions}. 

\begin{table}[htb]
\centering
    \begin{tabular}{cccc}
        \hline
$\lambda_{\rm c}~(\mu\rm m)$	&	$\lambda_{\rm Blue}$	&	$\lambda_{\rm Red}$	&	Median $\sigma_\mathcal{F}$ (ppm) \\
\hline
0.80317	&	0.79921	&	0.80712	&	374	\\
0.81107	&	0.80712	&	0.81502	&	348	\\
0.81898	&	0.81502	&	0.82293	&	329	\\
0.8506	&	0.82293	&	0.87826	&	102	\\
0.88485	&	0.87826	&	0.89144	&	177	\\
0.89473	&	0.89144	&	0.89803	&	241	\\
0.90791	&	0.89803	&	0.91779	&	133	\\
0.92168	&	0.91779	&	0.92558	&	201	\\
0.93307	&	0.92558	&	0.94057	&	142	\\
0.94436	&	0.94057	&	0.94816	&	195	\\
0.95955	&	0.94816	&	0.97094	&	109	\\
0.97474	&	0.97094	&	0.97853	&	185	\\
0.98233	&	0.97853	&	0.98613	&	183	\\
0.99941	&	0.98613	&	1.01270	&	97	\\
1.01713	&	1.01270	&	1.02156	&	164	\\
1.03042	&	1.02156	&	1.03928	&	114	\\
1.04307	&	1.03928	&	1.04687	&	173	\\
1.05067	&	1.04687	&	1.05446	&	173	\\
1.05826	&	1.05446	&	1.06206	&	173	\\
1.06543	&	1.06206	&	1.06881	&	183	\\
1.08568	&	1.06881	&	1.10255	&	81	\\
1.10593	&	1.10255	&	1.10930	&	181	\\
1.11268	&	1.10930	&	1.11605	&	182	\\
1.12323	&	1.11605	&	1.13040	&	125	\\
\midrule								
1.15882	&	1.14064	&	1.17700	&	123	\\
1.18555	&	1.17700	&	1.19410	&	172	\\
1.20265	&	1.19410	&	1.21120	&	169	\\
1.22010	&	1.21120	&	1.22900	&	162	\\
1.23750	&	1.22900	&	1.24600	&	164	\\
1.25545	&	1.24600	&	1.26490	&	154	\\
1.29543	&	1.26490	&	1.32596	&	84	\\
1.33740	&	1.32596	&	1.34885	&	134	\\
1.36983	&	1.34885	&	1.39082	&	100	\\
1.40704	&	1.39082	&	1.42325	&	114	\\
1.43020	&	1.42325	&	1.43715	&	174	\\
1.46032	&	1.43715	&	1.48348	&	96	\\
1.50896	&	1.48348	&	1.53444	&	91	\\
1.54371	&	1.53444	&	1.55297	&	151	\\
1.56073	&	1.55297	&	1.56850	&	166	\\
1.58050	&	1.56850	&	1.59250	&	134	\\
1.60450	&	1.59250	&	1.61650	&	134	\\
1.62850	&	1.61650	&	1.64050	&	136	\\
\hline
    \end{tabular}
    \caption{Central wavelengths, bin edges, and uncertainty information for the WFC3 light curves analyzed in this paper. Rows above the split correspond to S22/G102 and below to F21/G141.}
    \label{tab:binned-lcs}
\end{table}

\begin{figure*}[ht!]
\centering
    \includegraphics[width=0.93\textwidth]{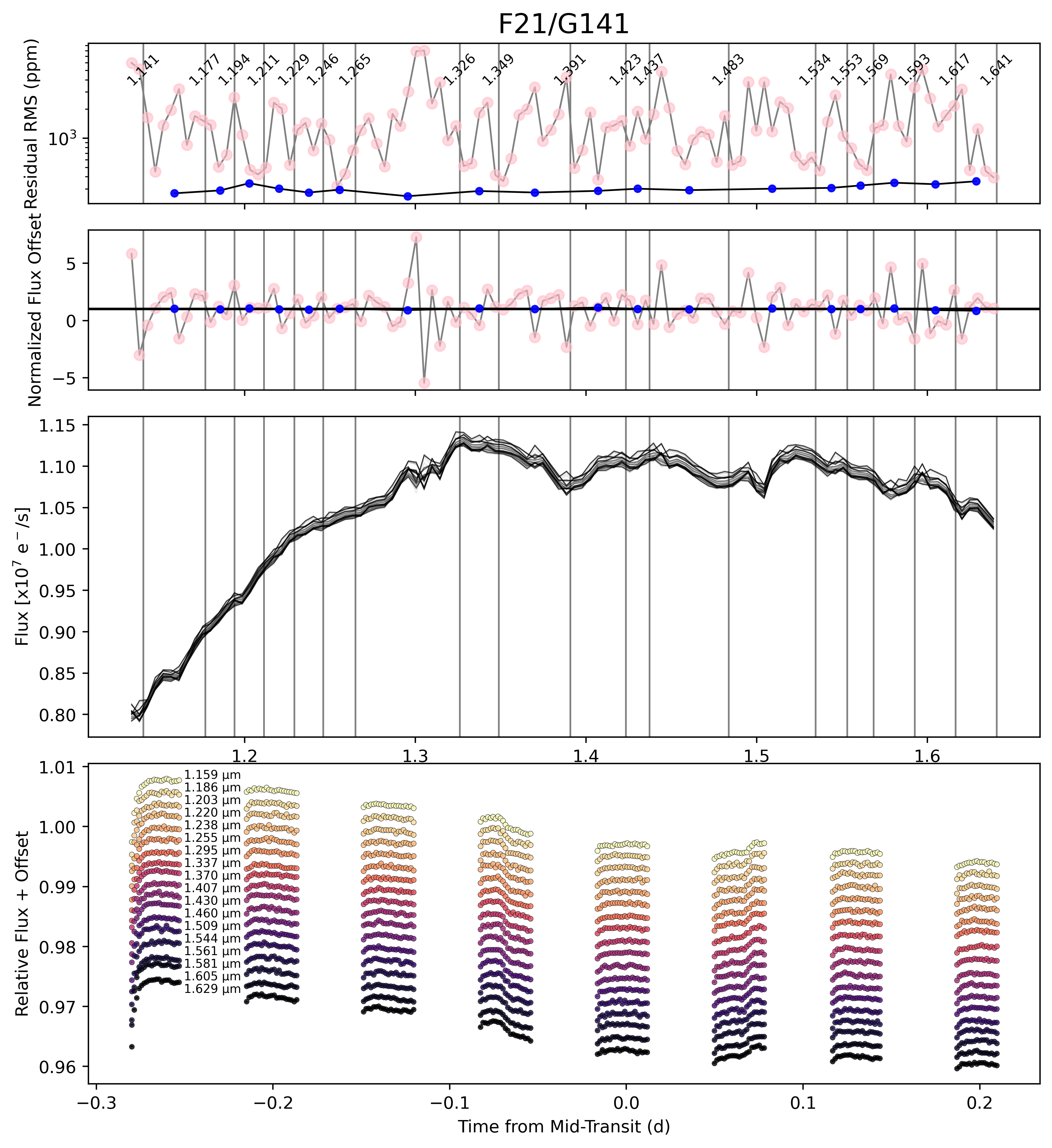}
\caption{Binning scheme for F21/G141 with vertical red bars to denote the bin edges.
Top row: the standard deviation of the residual RMS calculated for a linear model fit to that wavelength's light curve, both for the unbinned light curves (pink) and the resulting binned light curves (blue). 
Second row: The median-flux offset between orbit 2 and orbit 1 for each instrument-resolution (pink) and binned (blue) light curve, normalized to the equivalent calculation in the bandpass-integrated light curve. 
Third row: time-series spectra (unbinned in time or wavelength) overplotted to demonstrate how much variation there is in time in the observed spectra. 
Bottom row: the binned light curves used in this analysis.}   
\label{fig:F21binning}
\end{figure*}

\begin{figure*}[ht!]
\centering
    \includegraphics[width=0.93\textwidth]{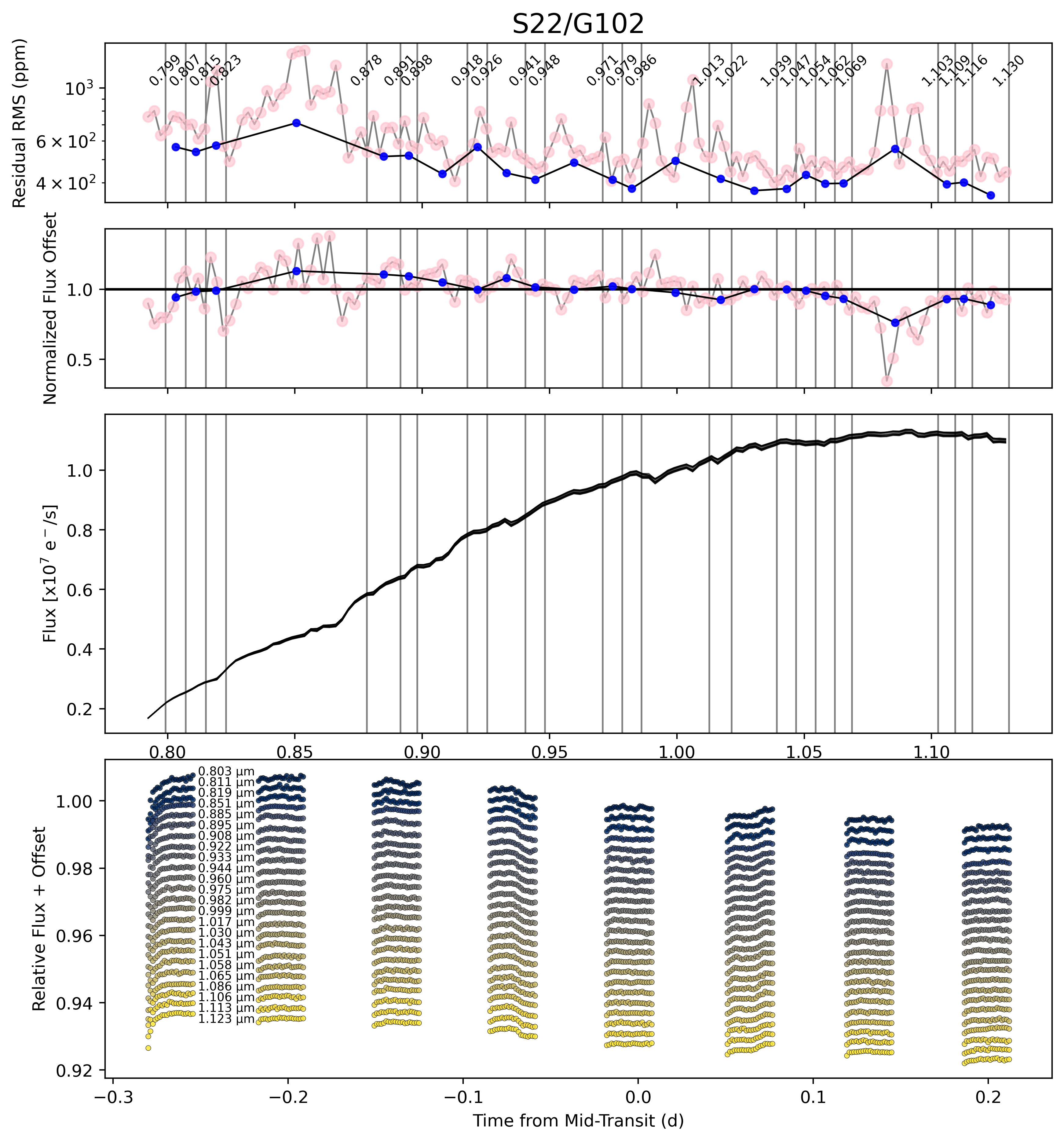}
\caption{Same as Figure \ref{fig:F21binning} for the S22/G102 bandpass.}   
\label{fig:S22binning}
\end{figure*}

\section{Observations and Data Processing} \label{observations_and_data}

In this section, we introduce our HST WFC3 observations in \S\ref{subsection:data} and summarize data reduction using PACMAN in \S\ref{subsection:processing}. Unstable pointing requires that we carefully consider how to bin our data in wavelength in order to produce spectroscopic lightcurves, which we discuss in \S\ref{subsection:binning-lcs}.

\subsection{Hubble Space Telescope Observations} \label{subsection:data}
We observed two transits of AU Mic b using HST (GO 15836; PI Newton). Transits were observed on 2021 August 30 using the G141 ($\sim$1.1-1.7 $\mu$m) grism (hereafter F21/G141) and 2022 April 14 using the G102 ($\sim$0.78-1.15 $\mu$m) grism (hereafter S22/G102). Details can be found in Table \ref{tab:1-observations}.
All data were collected in ``round-trip'' mode that includes forward and reverse scanning \citep[e.g., ][]{Deming2013}. Because the readout flux is offset between scan directions \citep{Knutson2014a}, we end up with a total of two distinct time series spectra of this target in either visit. 
F21/G141 observations had exposure times of 4.97 seconds, covering 140 pixels (scan rate of 28 pix/s, or 3.6"/s) on the 256$\times$256 subarray and totaling 324 exposures over 8 HST orbits.
S22/G102 observations had exposure times of 9.66 seconds, covering 137 pixels (scan rate of 14 pix/s = 1.84”/s) on the 256$\times$256 subarray, totaling 320 exposures over 8 HST orbits.
While in both cases the flux of the star is very high ($\sim$45k counts), we examined the difference images of several exposures in the first and second orbits where flux is highest and found no clear signs of saturation or non-linearity.

\begin{table*}[htb]
\centering
\begin{tabular}{lcc}
    \hline
    \textbf{Parameter Name} & \textbf{Value} & \textbf{Citation} \\ \hline
    R$_{\rm s}$ (M$_{\odot}$) & 0.80 & \citet{Gallenne2022, Donati2023, Waalkes2024}\\ 
    M$_{\rm s}$ (M$_{\odot}$) & 0.60 & \citet{Gallenne2022, Donati2023}\\ 
    $\rm [Fe/H]$ & 0.0 & \\
    log \textit{g} & 4.5 & \citet{Donati2023} \\
    P$_{\rm rot}$ (days) & 4.86 & \citet{Martioli2021,Donati2023} \\
    \hline
    P$_{\rm orb}$ (days) & 8.463 & \citet{Wittrock2023,Donati2023} \\
    \textit{i} ($^\circ$) & 89.5 & \citet{Klein2021,Martioli2021} \\
     &  & \citet{Gilbert2022,Wittrock2023} \\
    \textit{e} & 0 & \citet{Donati2025} \\

    \hline
\end{tabular}
\caption{System parameters fixed in this study.}
\label{tab:system_params}
\end{table*}

\subsection{Data Reduction and Processing} \label{subsection:processing}

To reduce the WFC3 FITS files we use the PACMAN \citep{ZiebaKriedberg2022} pipeline built specifically for handling WFC3 transmission data.
PACMAN reads the ima.fits files (Stage 00\footnote{https://pacmandocs.readthedocs.io/en/latest/}), downloads the HST ephemerides for the visit(s) (Stage 01), and then performs a barycentric correction to convert MJD (the default unit in the header) to BJD (Stage 02). 
The pipeline then downloads a template stellar spectrum \citep[which we specified to be from ][]{Kurucz1993} and multiplies it by the grism throughput to create a reference spectrum (Stage 03) for wavelength calibration.
The F21/G141 observations were accompanied by an image taken with the F164N filter, which is used for wavelength calibration.
However, due to an error in telescope operations, the star did not fall on the detector with this setup.
Without the star on the direct image, PACMAN's automated extraction routines fail in Stage 10 (fitting for the location of the star) so we manually specify the location of the star by giving pixel locations that do not physically exist to ``trick'' the pipeline into finding the spectrum and performing optimal extraction (Stage 20).
This issue is not present in the S22/G102 visit.

The optimal extraction routine is based on the algorithm presented in \citet{Horne1986}, which performs sky subtraction with a weighted-least-squares fit and masks cosmic rays based on their distortion effects in the spatial direction, thereby avoiding the accidental masking of emission lines.
PACMAN determines the aperture width for optimal extraction by finding the rows that have the largest change in median flux across each column of the 2D spectral files, looping through all up-the-ramp samples to create a single extracted spectrum for each exposure.
We perform our own binning and modeling of the spectroscopic light curves following completion of Stage 20 of the PACMAN pipeline (see Section \ref{subsection:binning-lcs}). 
While PACMAN performs a wavelength calibration in Stage 20, we perform an additional correction on each extracted light curve.
To do this, we calculated a median spectrum from the extracted light curves and fit for a wavelength shift and stretch for each spectrum in time against this template.

We trim regions near the steep drop in grism sensitivity, resulting in 138 pixel-level light curves spanning 0.79-1.13 $\mu$m (S22/G102) and 110 pixel-level light curves spanning 1.14-1.64 $\mu$m (F21/G141).
Figure \ref{fig:untrimmed_data} shows the median spectra and white light curves produced by this process.
In each light curve we see evidence of astrophysical signals from the star (rotation, spot occultations, and flares), planet (transit), as well as the well-known HST ramp and breathing effects \citep{Wakeford2016}.
Some correlated noise is also expected in the pixel-level light curves as the resolution element of HST is slightly larger than 1 pixel. 

We bin both scan directions into a single spectroscopic light curve.
We sum the corresponding fluxes from both scan directions and calculate uncertainties in quadrature from the photon uncertainties in both directions.
The observation times and wavelengths differ between the two scan directions so we take the mean of the time and wavelength arrays and finally normalize the combined light curves in each wavelength by their mean.
Combining data from both scan directions increases signal-to-noise under the assumption that the systematics are shared across scan directions and allows us to average out the flux offsets we see between scan directions in the F21/G141 data (Figure \ref{fig:untrimmed_data}).

We trim all observations from HST's first orbital visit and the first observation from each subsequent orbital visit. This excludes observations known to be more heavily affected by systematics, and is a standard practice in the analysis of HST/WFC3 time series data \citep{Wakeford2016}.
In addition, the third HST orbit in our S22/G102 observations is noticeably impacted by a stellar flare affecting $\geq50\%$ of the 45-min orbit so we trim this entire orbit from the data. 
The F21/G141 data do not show any obvious flares so we do not trim any orbits on this basis. 
The lack of obvious flares outside of S22/G102 orbit 3 is of note because we expect frequent flares on this star \citep[][]{Feinstein2022a, Gilbert2022, Rockcliffe2023, Paudel2024}. Their absence implies that they are either confounded with other signals in the data or happened to occur in between HST orbits. In any event, this likely has unresolved effects on the results of this work. The fully processed broadband light curves have a median photon uncertainty of 40-50 ppm while individual light curves at the native time and wavelength resolutions have median photon uncertainties of 300-600 ppm.

\subsection{Spectroscopic lightcurves} \label{subsection:binning-lcs}

We bin the spectroscopic time-series data in wavelength to produce 24 light curves in the G102 bandpass and 18 light curves in the G141 bandpass.
The primary consideration in this step was in determining the optimal bin configuration to mitigate the variable PSF, which arose from an unstable scan along the detector.

\paragraph{Unstable Scanning} 
In a stroke of poor luck, both visits appear to have experienced an unstable scan similar to what is shown in Figure 11 of \citet{Sing2018}.
Essentially, the telescope is shaking throughout the observations due to poor tracking, creating a wavelength- and time-variable PSF that cannot be corrected through the standard data reduction process.
This effect can be seen in the wave-like pattern in the 2D spectra shown in Figure \ref{fig:2Dspec} and our attempts to fix it are described below.

The effects of HST's unstable scan persist through the wavelength correction applied in the data processing stage and therefore some additional correction is necessary.
For a given wavelength column along the detector, the light curve baseline often varies orbit-to-orbit on a scale much larger than the scatter of the intra-orbit exposures. 
We expect correlated noise between adjacent pixel-level light curves due to HST's resolution element being larger than 1 pixel, but we found this issue to remain even when testing many different configurations of uniformly sized bins.
Furthermore, since this effect is highly correlated with wavelength (see panel 1 of Figures \ref{fig:F21binning} and \ref{fig:S22binning}) and not a Gaussian-distributed effect, simply re-scaling uncertainties on the pixel-level light curves does not address the issue, and we find that we can make the most of the data by carefully placing wavelength bins. 

\paragraph{Choosing Bins}
To determine where the bin edges should be placed, we fit a linear model to the OOT light curves and measured the RMS of the residuals from that fit.
While a straight line is not a particularly good fit to the data, the variations away from this line are highly informative.
We use the RMS of the linear-fit residuals and visual inspection to select bins edges.
Along with the linear RMS, we calculate and inspect a metric defined as the difference in median fluxes between orbit 3 and orbit 2, where this effect is typically the most obvious.
For each spectroscopic light curve, this baseline flux difference is normalized by the equivalent calculation of the bandpass-integrated light curve.
Quantitatively, the Normalized Flux Offset (NFO) is calculated as $$\rm NFO=\frac{\left[Median(F_{Orbit~3})-Median(F_{Orbit~2})\right]_{\lambda_i}}{\left[Median(F_{Orbit~3})-Median(F_{Orbit~2})\right]_{white}}$$ for each of $i$ wavelengths.
This calculation is repeated on the binned light curves to examine the improvement from a given bin placement. 
In panel 2 of Figures \ref{fig:F21binning} and \ref{fig:S22binning}, points near 1 (horizontal black line) indicate an orbit-to-orbit offset that is similar to the integrated light curve offset between orbits, indicating a more or less ``unaffected" region of the light curve. On the contrary, points far from 1 represent extreme signatures of the effects of the variable PSF.

Our solution is ultimately imperfect and incentivizes a significant reduction in spectral resolution, but future analysis might make use of more complex deep-learning or gaussian process methods to correct for this and recover more precise light curves. 
Details on the final binned light curves can be found in Table \ref{tab:binned-lcs}.


\section{Methods} \label{methods}
In this section we first introduce the separate stages of analyzing the binned light curves in \S\ref{subsection:overview}. In \S\ref{subsection:spec-decomp} we describe the spectral decomposition model used to constrain stellar photospheric temperatures. We then go on to construct a systematics model in \S\ref{subsection:systematics-model} followed by a spotted stellar photosphere model in \S\ref{subsection:spotted-whitelight-model} which is used to identity spots on the transit chord. These spots are subsequently masked when we model the spectroscopic light curves and extract a contaminated planetary transmission spectrum, as outlined in \S\ref{subsection:speclc-model}. Details on the MCMC inference structure for the light curve fitting is presented in \S\ref{subsection:inference}. Finally, the methods used to perform atmospheric retrieval and model the TLS effect on the transmission spectrum are described in \S\ref{sec:methodretrieval}.

\subsection{Light Curve Analysis Overview} \label{subsection:overview}

Following previous best practices, we analyze our transit light curves by extracting instrument systematics and wavelength-independent transit parameters from the white light curve and then modeling the de-trended spectroscopic transits to extract transit depths at each wavelength. 
We treat contamination from unocculted active regions on the host star as a component in the measured transmission spectrum during atmospheric retrieval (see Section \ref{sec:methodretrieval}).
Clear signs of rotation and spot occultations in our spectroscopic WFC3 lightcurves likely warrant a more cautious approach. However, given the complex multi-variate physical reality of this system alongside these observations' HST systematics, we focus here on the standard analysis, divided into three steps to reduce dimensionality and convergence time:
\begin{enumerate}

    \item \textbf{Decomposition of the OOT stellar spectrum.} We model both visits together, with temperature components and filling factors shared between visits. This provides tight constraints on the stellar photosphere's spectral components, allowing us to fix the spot contrast in the white light curve analysis. These measurements also inform our priors for the atmospheric retrieval. 
    
    \item \textbf{Analysis of the broadband transit light curve and instrument systematics.} We model the broadband light curves separately by visit, allowing us to measure and fix achromatic systematic (ramp \& breathing) and transit ($\rm t_0~\&~a/R_*$) parameters for subsequent transit light curve modeling.
    We perform broadband analysis using two different models -- the first where spots are modeled on the star to account for rotation and occultations, and the second using a linear model.
    The spotted stellar model helps us identify features of the in-transit light curves which are likely due to spots and trim those regions for the transit modeling.
    The linear model is used for measuring and fixing the instrumental parameters for the spectroscopic light curve analysis. 
    
    \item \textbf{Contaminated transit.} We model the transit in each wavelength bin with a linear baseline in order to extract transmission spectra that are implicitly contaminated and will be analyzed with a joint planetary $+$ TLS model in the atmospheric retrieval stage.

\end{enumerate}

\subsection{Spectral Decomposition} \label{subsection:spec-decomp}

We use the spectroscopic light curves to derive the stellar SED, first masking the planetary transit and then calculating a single uncertainty-weighted stellar spectrum for either visit.
To perform spectral decomposition, we convert our raw spectrum (units of counts) into a physical spectral energy distribution (SED, units of flux per unit wavelength).
We divide each spectrum by the exposure time for the instrument and the instrumental bin width to obtain units of counts per second per nm.
We then divide by the first order instrumental sensitivity curves \citep{Kuntschner2011wfc..rept....5K} to convert counts per second to flux in a procedure similar to that described in \citet{Garcia2022,Narrett2024}.
There is uncertainty in the absolute flux calibration level of the sensitivity curve and we do not aim to revise measurements of the stellar radius or distance in this work, so we mean-normalize the calibrated spectrum to handle it in relative flux units.

\begin{table*}[tb]
    \centering
    \begin{tabular}{lccc}
        \hline
        \textbf{Parameter Name} & \textbf{Full Model} & \textbf{TLS-Only} & \textbf{Step Function} \\ \hline
        R$_{\rm s}$ (R$_{\odot}$) & 0.8 (fixed) & 0.8 (fixed) & 0.8 (fixed) \\ 
        R$_{\rm p}$ (R$_{\oplus}$) @ 1 bar & $\mathcal{U}(2,5)$ & $\mathcal{U}(2,5)$ & $\mathcal{U}(2,5)$ \\ 
        T$_{\rm phot}$ (K) & $\mathcal{N}(3891,37)$ & $\mathcal{N}(3891,37)$ & 3891 (fixed) \\
        T$_{\rm spot}$ (K) & $\mathcal{N}(3020,69)$ & $\mathcal{N}(3020,69)$ & N/A \\ 
        f$_{\rm spot}$ & $\mathcal{U}(0,1)$ & $\mathcal{U}(0,1)$ & N/A \\
        $\epsilon'$ & $\mathcal{U}(0,max(\epsilon_D))$ & $\mathcal{U}(0,max(\epsilon_D))$ & $\mathcal{U}(0,max(\epsilon_D))$ \\
        $\Delta D$ (ppm) & $\mathcal{U}(-1000,1000)$ & $\mathcal{U}(-1000,1000)$ & $\mathcal{U}(-1000,1000)$ \\
        M$_{\rm p}$ (M$_{\oplus}$) & $\mathcal{U}(1,50)$ & 50 (fixed)  & 50 (fixed) \\
        T$_{\rm atm}$ (K) & $\mathcal{U}(200,1000)$ & 200 (fixed) & 200 (fixed) \\
        ${\rm [M/H]}$ ($\times$ Solar) & $\mathcal{U}(-1,3)$ & -1 (fixed) & -1 (fixed) \\
        C/O & $\mathcal{U}(0.05,1.5)$ & 0.59 (fixed) & 0.59 (fixed) \\
        log(P$_{\rm cloud}$) (Pa) & $\mathcal{U}(-3.999,7)$ & -3.999 (fixed) & -3.999 (fixed) \\
        log($f_{\rm scatt}$) & $\mathcal{U}(0,5)$ & 0 (fixed) & 0 (fixed) \\
        $m_{\rm scatt}$ & $\mathcal{U}(0,20)$ & 4 (fixed) & 4 (fixed) \\
        \hline
    \end{tabular}
    \caption{Priors for the three \texttt{PLATON} retrievals, where R$_{\rm p}$ is the planet radius, $\epsilon'$ is the additional uncertainty on the transit depth per wavelength bin added in quadrature to the actual uncertainty $\epsilon_D$, $\Delta D$ is the transit depth offset between the two observations, T$_{\rm atm}$ is the isothermal temperature of the model atmosphere, log(P$_{\rm cloud}$) is the base-10 log of the cloud top pressure, log($f_{\rm scatt}$) is the base-10 log of the multiplicative enhancement factor on the optical scattering slope, and $m_{\rm scatt}$ is the slope of the scattering itself, where 4 denotes Rayleigh scattering.}
    \label{tab:platon_priors}
\end{table*}

Spectral decomposition entails calculating a weighted combination of spectra of \textit{i} different temperatures:
\begin{equation}
\mathrm{S_{\lambda}}=\sum_i f_i S(\lambda, \mathrm{T_i}),
\end{equation}
where $S_{\lambda}$ is the model SED, $f_i$ is the filling factor for a given flux component, and $S(\lambda, \mathrm{T}_i)$ is the (interpolated) model spectrum for a given temperature.

More than one temperature component is needed to produce the observed AU Mic stellar spectrum \citep[see, e.g., ][]{Gilbert2022,Donati2023,Waalkes2024,Donati2025}, but no evidence has yet been found for $>3$, so we explored the presence of two or three temperatures under a variety of different initial assumptions and prior boundaries.

We use the BT-Settl AGSS2009 spectral library \citep{Allard2012BT} with fixed metallicity (=0) and log(g) (=4.5) ranging in temperature of 2300-10000 K at 100 K intervals.
We perform a linear interpolation between models for sampled temperatures that fall between adjacent gridpoints. 
To jointly model the spectra of both visits, we include an additive term, $b_{G102}$, which shifts the S22/G102 spectrum along the flux axis to match the F21/G141 spectrum. Slope parameters are also included to correct for linear wavelength-dependence along the detector. 
The combined model SED is subsequently mean-normalized and smoothed with a Gaussian kernel before comparison to the data. 

\subsection{Systematics Model} \label{subsection:systematics-model}
HST observations with the WFC3 instrument are subject to an intra-visit exponential ramp effect and a breathing effect which creates changes on the timescale of HST's orbital phase \citep[see ][ for a review of WFC3 systematics]{Wakeford2016}.
Modeling the white-light curve allows us to measure the parameters of these instrumental effects, which will then be fixed for the spectroscopic light curve fitting.

To determine the best set of systematics models for our data, we ran both visits through the ExoTiC-ISM \citep{Wakeford2016,Laginja2020} software.
This software runs a marginalization routine across 49 different model combinations, fitting for the exponential ramp as well as a visit-average slope and polynomial functions for the HST breathing effect and wavelength shift on the detector.
For our data, the marginalization routine indicated that a model involving the ramp effect as well as 4th order polynomials for breathing and wavelength-shift systematics was the best choice.
We adopt this parameterization, with the exclusion of a wavelength-shift systematic, as this systematic is addressed in the PACMAN pipeline.
Our systematics model is therefore a function of HST orbital phase ($\phi$) and the intra-orbit ramp phase ($\psi$) following the form of \citet{Stevenson2014a}. Our systematics model takes the form:
\begin{equation}
    S(t, \lambda) =  [1+\sum^{4}_{j=1}b_j\phi^j] \times [1 - e^{r_1 \psi + r_2} + r_3 \psi] 
\end{equation}
where $\sum^{4}_{j=1}b_j\phi^j$ is the breathing effect polynomial in HST orbital phase (P$_{\rm orb, HST} = 0.066$d) and $\psi=(t-t_{0,j})/P_{\rm HST}$ represents the phase over which the exponential ramp occurs, calculated relative to the first exposure of each orbit ($t_{0,j}$).

\subsection{White Light Model} \label{subsection:spotted-whitelight-model}
We use two distinct models for the white light curve analysis - the first model uses \fleck, a JAX-based forward model for transits of exoplanets on rotating, spotted stars \citep{Morris2020a,Morris2020b}.
In the second case, we trim the spotted regions of the transit chord identified via \fleck analysis and fit the light curve with a linear baseline.

\fleck models spots as circular projections on the stellar surface, parameterized by spot and photosphere temperatures (T$_{\rm spot}$ and T$_{\rm phot}$), radius (R$_{\rm spot}$ in units of stellar radii), latitude, and longitude.
The stellar photosphere is parameterized with two spectral temperatures, T$_{\rm phot}$ for the ambient photosphere and T$_{\rm spot}$ for the spots, with values fixed based on results from the OOT SED model.
We fix the stellar rotation period to 4.86 days and stellar inclination to 90$^\circ$.

After trimming points associated with spot occultations in the mid-transit and egress orbits for both visits, we model the white light curves again with a non-astrophysical baseline parameterized by a single term for the visit-long slope. 

The transit model is computed within \fleck based on \citet{Agol2020}. 
For both the spotted and linear baseline models, we fit the transit with mid-transit time ($t_0$), quadratic limb darkening coefficients ($u_1,u_2$), and transit depth ($r_p/r_*$) as free parameters, along with the instrumental systematic parameters.
We fix the planet orbital period to 8.463 days, eccentricity to 0, and inclination to 89.5$^\circ$.
We included the scaled semi-major axis ($a/R_*$) as a free parameter in early model iterations and fix it to 19.15 for the final analysis.

\subsection{Spectroscopic Light Curve Model} \label{subsection:speclc-model}
Spectroscopic light curves are modeled with a non-astrophysical (i.e., linear baseline) approach similar to standard HST transit analysis.
We use the results of the \textit{spotted} white-light modeling only to identify and trim spot occultations from the transit chord.
The results from the \textit{linear} white light curve fit are used to fix the instrumental parameters for modeling the spectroscopic light curves.
The spots we trim from the transit chord account for over half of the in-transit orbit and the first third of the egress orbit.

With instrumental and achromatic transit parameters fixed based on the white light curve models, we fit the spectroscopic light curves independently in wavelength by fitting for a time-dependent linear baseline term, planetary radius ($r_p/r_*$), error inflation term (10$^\beta$), and quadratic limb-darkening terms ($u_1$, $u_2$).

\subsection{Light Curve Model Inference} \label{subsection:inference}
We perform an MCMC inference when fitting the OOT SED, white light curves, and spectroscopic light curves.
This inference is built on JAX -- a library designed for scalable, hardware-accelerated linear algebra on CPU, GPU, or T(ensor)PU with automatic differentiation.
Autodifferentiation provides the gradient of the likelihood with respect to each free parameter, enabling more efficient sampling of complex and multi-modal posterior distributions. 

In this analysis we perform hierarchical Bayesian Modeling using the No U-Turns Sampler \citep[NUTS, ][]{HoffmanGelman2011}.
NUTS is a variation of Hamiltonian Monte Carlo (HMC) designed for exploring multi-model and highly co-variant posteriors. The walkers progress through the likelihood space, without quickly returning to regions of low- or zero-probability, which efficiently generates independent samples.
We draw samples with NUTS using NumPyro \citep{phan2019composable} which provides a front-end to the Pyro \citep{bingham2019pyro} probabilistic programming library.

In the NUTS/NumPyro setup, each model parameter has an associated Gelman-Rubin statistic \citep[\^{r};][]{Gelman1992}, allowing us to directly compare how well the posterior distributions are being sampled for each parameter. 
As chains converge, they approach $\hat{r}\rightarrow 1$ from above.
In practice, several iterations of each model's inference were required in order to minimize spurious solutions caused by chains getting stuck in local minima, often due in part to the systematic issues in these data. 

For the white light fits we use  60000 steps with a 6000-step burn-in per chain on 12 parallel chains.
Spectroscopic curves were modeled independently with 4000 steps with a 2000-step burn-in per chain over 8 chains, resulting in 32000 samples.

\begin{table}[htb]
    \centering
\begin{tabular}{lccc}		
\toprule			
Parameter &  1-T Model & 2-T Model   &   3-T Model  \\
T$_{\rm phot}$ [K] & $3779\pm19$ & $3944\pm32$ & $3891\pm37$\\
T$_{\rm spot}$ [K] &  & $2930\pm69$ & $3020\pm69$\\
T$_{\rm other}$ [K] &  &  & $6163\pm412$\\
$f_{\rm phot}$	& 1.0 & $0.74\pm0.03$ & $0.64\pm0.05$\\
$f_{\rm spot}$	&  & $0.26\pm0.03$ & $0.33\pm0.05$\\
$f_{\rm other}$	&  & & $0.03\pm0.01$ \\
$\Delta$ELPD & $37.1\pm8.9$ & $18.4\pm5.4$ & 0 \\
Weight & 0.000004 & 0.002494 & 0.997501 \\
\bottomrule	
\end{tabular}
\caption{Median $\pm$ 1-$\sigma$ measurements for the temperature and filling factors measured by the joint-visit SED model. In the last two rows are statistics for model comparison: The difference (relative to the best model) in the Expected Log-Predictive Density ($\Delta$ELPD) and the associated weights of each model.}
\label{tab:SED-results}
\end{table}

\begin{figure*}[ht!]
\centering
    \includegraphics[width=0.97\textwidth]{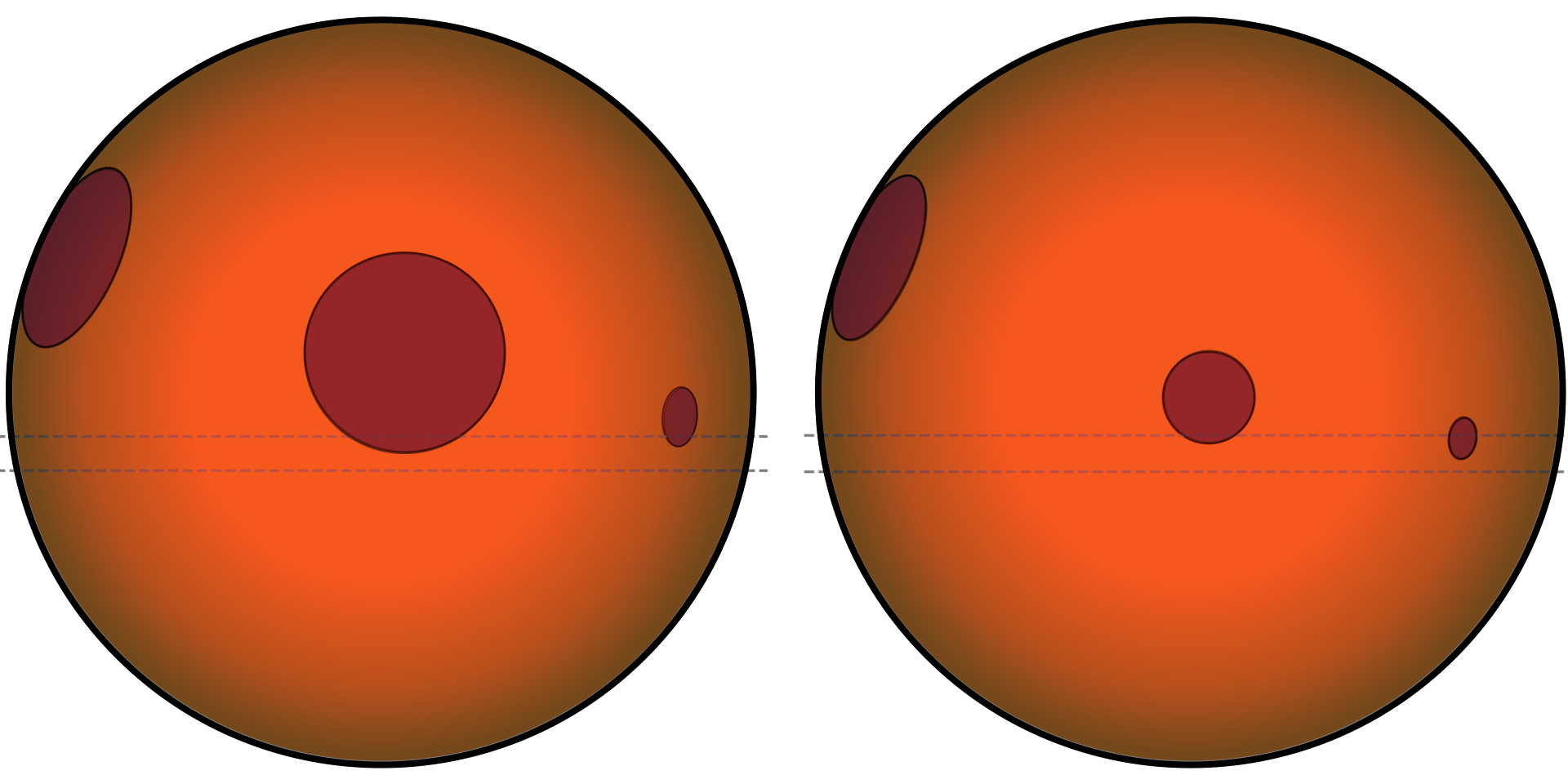} 
\caption{Stellar photosphere map (left: F21/G141, right: S22/G102) with associated active region locations and sizes measured by the astrophysical white light curve model. These models of the surface of AU Mic are simplifications based on a single bandpass and do not illustrate the complex reality of spot morphology. Rather, the locations and sizes of these spots represent the simplest configuration we explored which could describe apparent spot occultations and rotational modulation of the data.}
\label{fig:star_map}
\end{figure*}

\begin{figure*}[ht!]
    \includegraphics[width=0.47\textwidth]{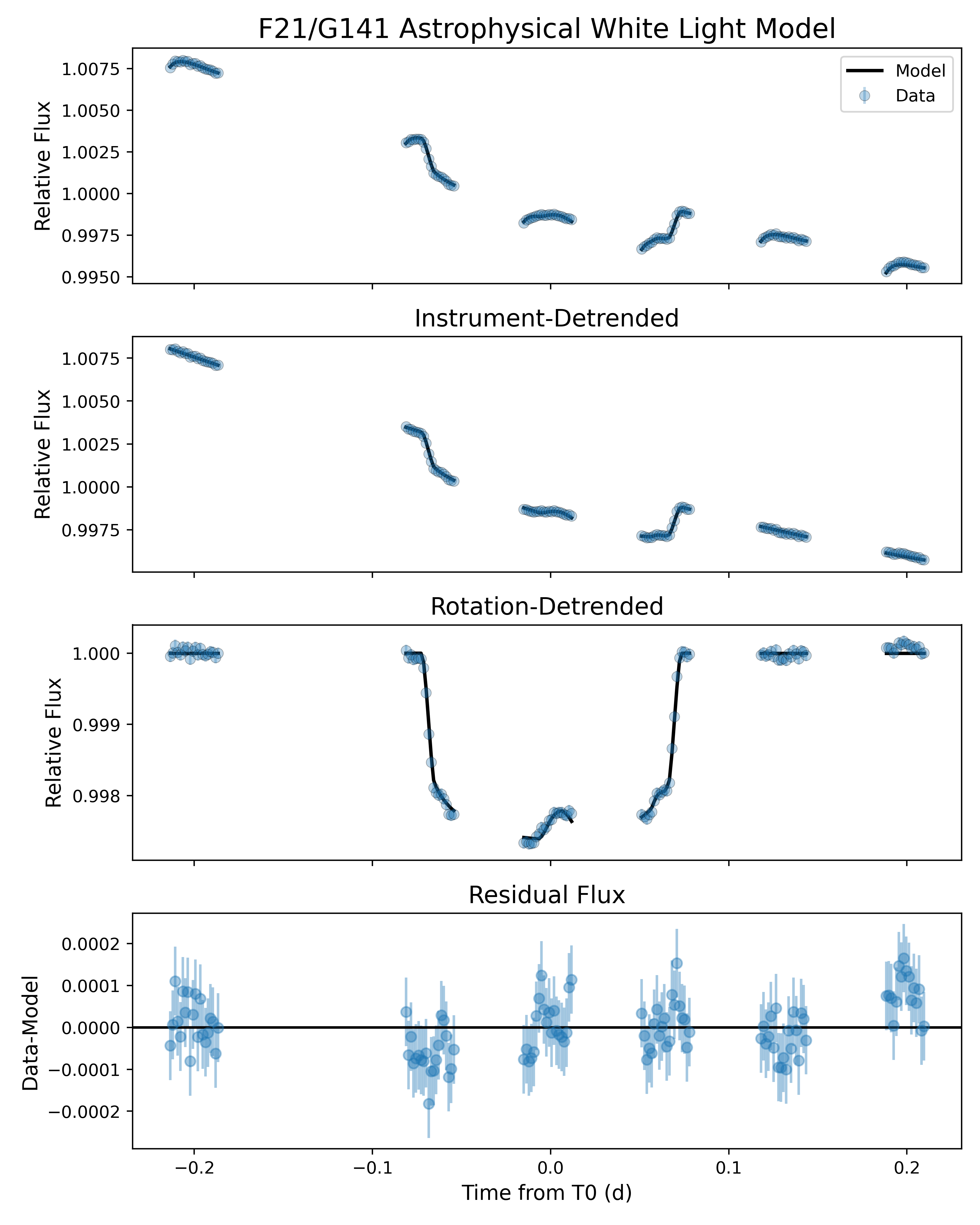}
    \includegraphics[width=0.47\textwidth]{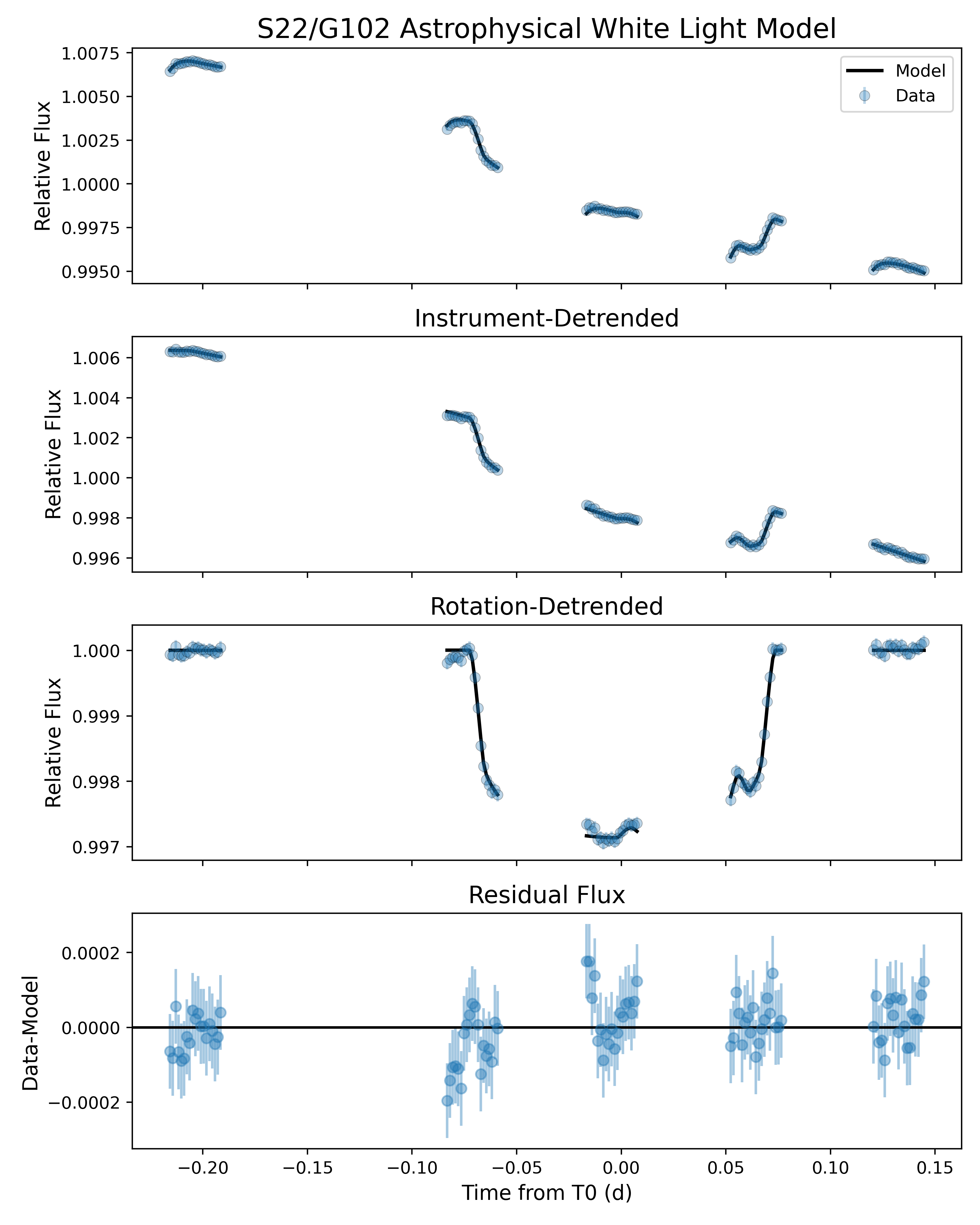}
\caption{Spotted white-light fits for F21/G141 (left) and S22/G102 (right). Models shown include both astrophysical and instrumental signals. Data are plotted with scaled uncertainties.
Spot occultations can be seen more clearly in the 3$^{\rm rd}$ row, where the light curves have had both the instrumental and rotational signatures removed. Significant structure in the residuals, especially the ingress and mid-transit of S22, indicate that some sources of astrophysical and/or instrumental variability remains unresolved. The beginning of the mid-transit orbit in S22/G102 has noticeably increased flux, but we choose not to fit this as a spot due to the limited number of points and large uncertainty in the source of the brightening.}
\label{fig:white-light-spotted}
\end{figure*}

\begin{figure}[ht!]
    \includegraphics[width=0.47\textwidth]{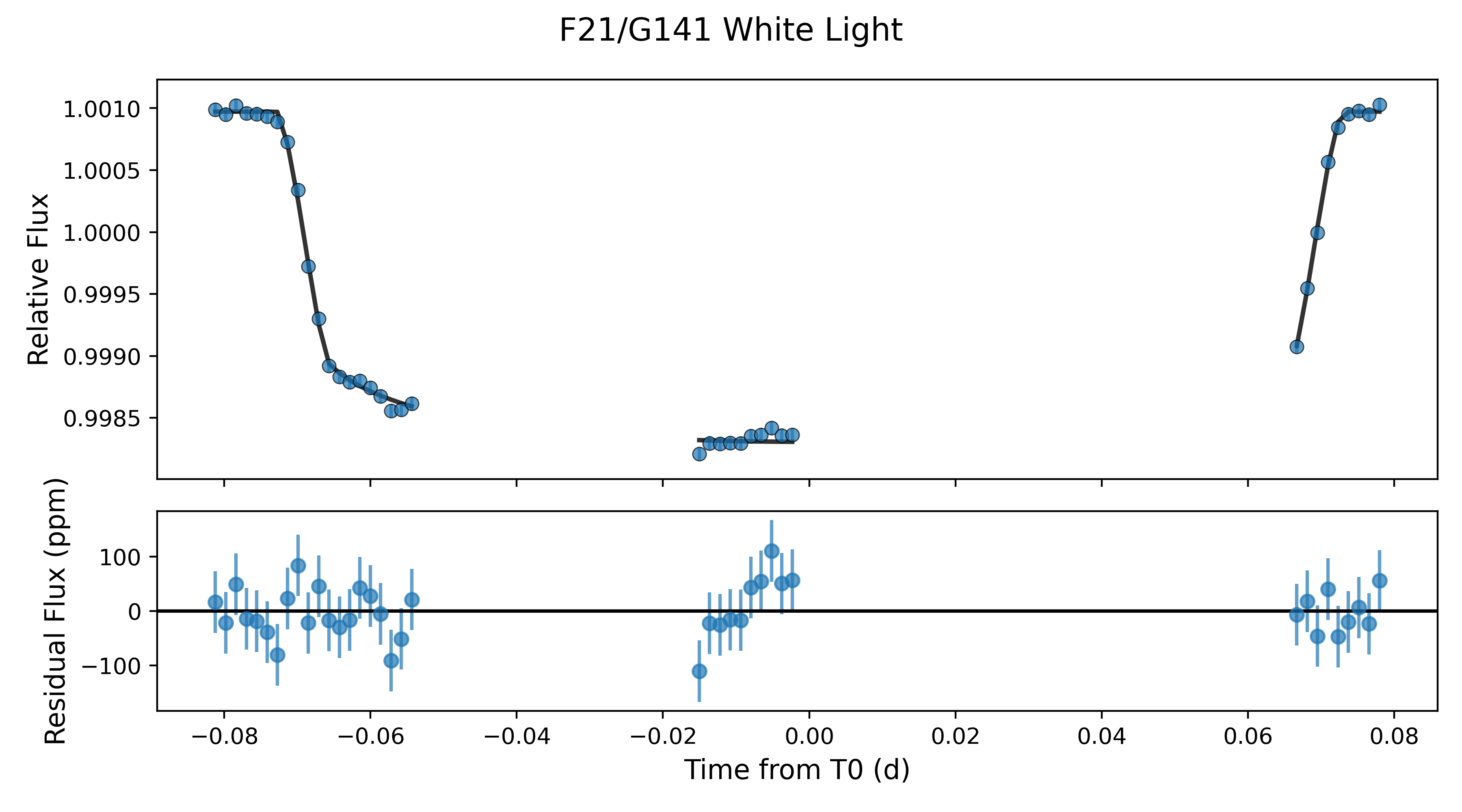} \\
    \includegraphics[width=0.47\textwidth]{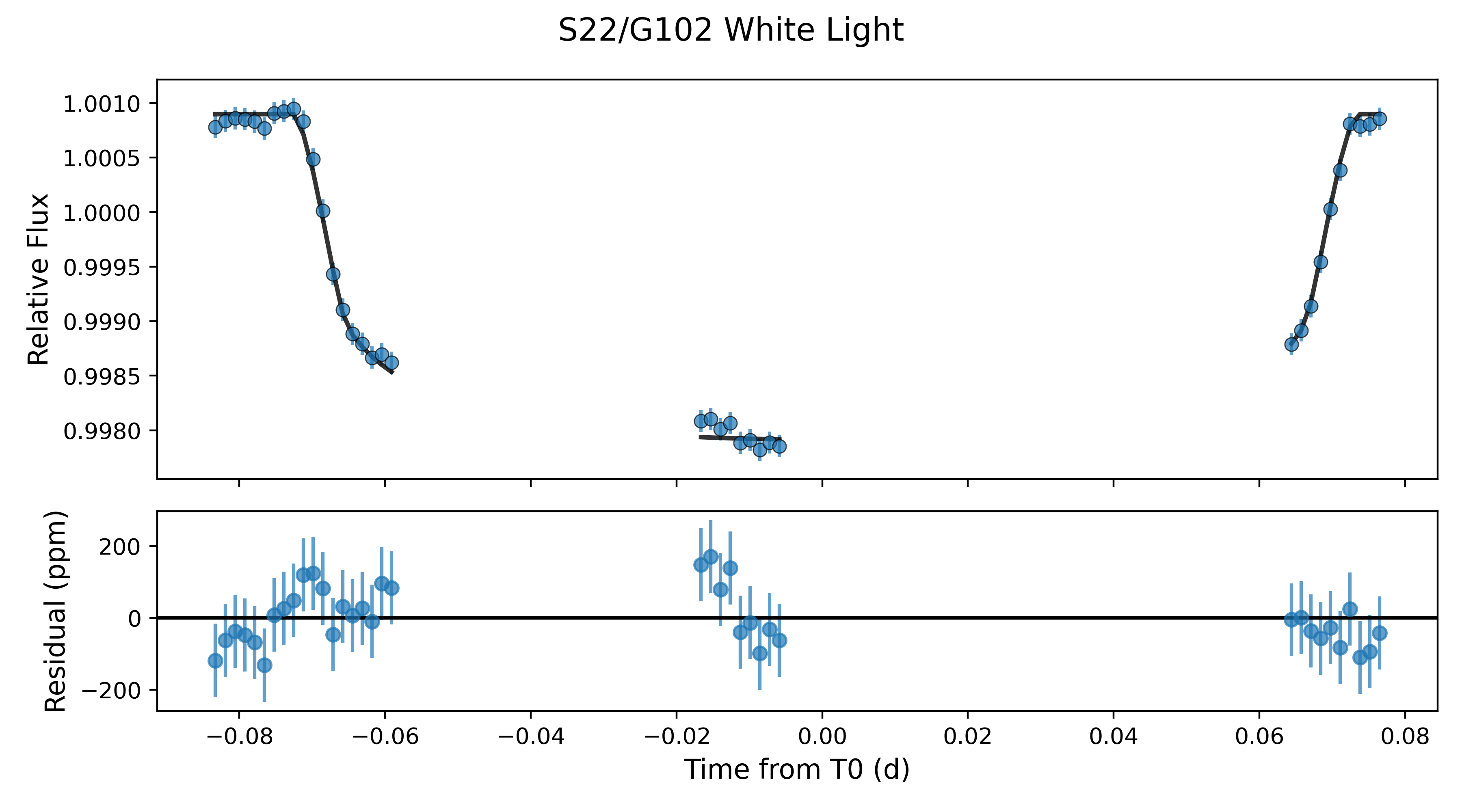}
\caption{Linear white-light curve models for F21/G141 (top) and S22/G102 (bottom). Here, the data have been normalized by the median instrumental model, and spot occultations have been trimmed from the transit chord.}
\label{fig:white-light-linear}
\end{figure}

\subsection{Atmospheric Retrieval}\label{sec:methodretrieval}

We conduct Bayesian retrievals on the finalized transmission spectrum using \texttt{PLATON}\footnote{https://github.com/ideasrule/platon} \citep[PLanetary Atmospheric Tool for Observer Noobs;][]{Zhang2019PLATON,Zhang2025} to infer the atmospheric composition of AU Mic b. \texttt{PLATON} is an open source retrieval code capable of retrievals across optical and IR wavelengths for transmission and emission observations of exoplanets. We use the default R=20,000 resampled opacities and the \texttt{pymultinest} sampler with 1,000 live points. 

Here we consider thermochemical equilibrium retrievals with the addition of the TLS effect. We do not consider free retrievals due to the limited number of data points and the SNR of the data. For the full list of chemical species \texttt{PLATON} treats in thermochemical equilibrium and the references for their opacities, we refer the reader to Table 2 of \citet{Zhang2025}. \texttt{PLATON} uses \texttt{FastChem}\footnote{https://github.com/NewStrangeWorlds/FastChem} \citep{fastchem1,fastchem2,fastchem3} to compute thermochemical equilibrium abundances and assumes the solar abundances of \citet{Asplund2021}. The TLS effect is parameterized in \texttt{PLATON} with a stellar photospheric temperature ($T_{\rm phot}$), a spot temperature ($T_{\rm spot}$) and a spot filling factor ($f_{\rm spot}$), related through 

\begin{equation}
    D_{\lambda}' = D_{\lambda}\frac{S(\lambda,T_{\rm phot})}{f_{\rm spot}S(\lambda,T_{\rm spot})+(1-f_{\rm spot})S(\lambda,T_{\rm phot})} 
\end{equation}

\noindent where $D_{\lambda}'$ and $D_{\lambda}$ are the TLS-impacted and original (atmosphere-only) transit depths, respectively. The stellar spectra $S(\lambda, \rm T_i)$ in PLATON are obtained from interpolating over the BT-NextGen (AGSS2009) stellar spectral grid \citep{Allard2012BT}, assuming logg = 4.5 and solar metallicity \citep{Zhang2019PLATON}. 

Table \ref{tab:platon_priors} presents the priors for the spectral retrievals. We consider three models for the transmission spectrum to assess what information can be extracted. The ``Full Model'' case assumes both an atmospheric component and a TLS component. The atmospheric component is determined by the atmospheric metallicity ([M/H]), C/O, and temperature (T$_{\rm atm}$), which together set the chemical composition assuming thermochemical equilibrium. We assume an isothermal pressure-temperature profile set to T$_{\rm atm}$ with a wide uniform prior around the estimated equilibrium temperature of AU Mic b ($\sim$600 K). We also include a gray cloud deck parameterized by a cloud top pressure P$_{\rm cloud}$, and a ``haze'' parameterized by an enhancement factor $f_{\rm scatt}$ on the optical Rayleigh scattering and a scattering slope $m_{\rm scatt}$, where Rayleigh scattering itself gives a slope of 4. The gravity of the planet, which controls the atmospheric scale height and therefore the amplitude of spectral features, is determined by the mass M$_{\rm p}$ and radius R$_{\rm p}$ of the planet. We consider wide uniform priors on both M$_{\rm p}$ and R$_{\rm p}$: for M$_{\rm p}$ the bounds exceed the range of values from the literature \citep{Donati2025,Cale2021}, while for R$_{\rm p}$, which is fixed to that at 1 bar for \texttt{PLATON}, we use wide bounds around the measured value from transits, as we do not know \textit{a priori} what pressures transits probe. The TLS component is parameterized by T$_{\rm phot}$, T$_{\rm spot}$, and $f_{\rm spot}$ as previously discussed. We use the T$_{\rm phot}$ and T$_{\rm spot}$ values derived here (see Section \ref{subsection:results-sed}), specifically from the 3-T model, but a wide prior for $f_{\rm spot}$ to investigate whether the transmission spectrum can provide consistent constraints. In addition, we include an error inflation term $\epsilon'$ that is added in quadrature to the existing errors of each wavelength bin, as well as an offset $\Delta D$ between F21/G141 and S22/G102, both of which have uniform priors; the prior on $\epsilon'$ is limited by the maximum error per bin of the actual data while the $\Delta D$ is given a wide prior to account for unknown systematics.  

In addition to the ``Full Model'' case, we also consider a ``TLS-Only'' case where the atmospheric transmission spectrum is assumed to be flat, with all features in the observed spectrum to be due to TLS. 
This allows us to investigate whether any constraints on AU Mic b's atmosphere are possible given the intense host star activity and the corresponding TLS effect.
To accomplish this, we fix P$_{\rm cloud}$ and [M/H] to be as low as possible in \texttt{PLATON} ($\geq$1 nbar and -1, or 0.1 $\times$ solar, respectively) to generate an extremely high cloud deck with few gaseous molecular absorbers, resulting in flat spectra. We also minimize T$_{\rm atm}$ and maximize M$_{\rm p}$ while fixing other atmospheric component parameters (C/O, $f_{\rm scatt}$, $m_{\rm scatt}$) to nominal values (see Table \ref{tab:platon_priors}), such that they do not introduce spectral features into the flat spectra. The TLS (T$_{\rm phot}$, T$_{\rm spot}$, $f_{\rm spot}$) and data ($\epsilon'$, $\Delta D$) parameters, as well as R$_{\rm p}$, are allowed to vary, with the same priors as in the ``Full Model'' case. Finally, we consider an even simpler ``Step Function'' case where we remove the TLS parameters such that the fitted model is just a flat line with an offset between the two datasets, leading to only three varied parameters (R$_{\rm p}$, $\epsilon'$, $\Delta D$). We include this case to assess whether the data can be explained by a completely featureless spectrum with neither TLS nor atmospheric information. 

\section{Results} \label{results}
Here we present our results starting with the temperature constraints from the OOT SED in \S\ref{subsection:results-sed} and followed by what we learned about the stellar surface from modeling the bandpass-integrated white light curves in \S\ref{subsection:results-whitelight}. The transmission spectrum results are described in \S\ref{subsection:results-transmission} and measurements of stellar contamination and the planet's atmosphere atmospheric retrieval are presented in \S\ref{subsection:results-retrieval}

\subsection{OOT SED}\label{subsection:results-sed}

Table \ref{tab:SED-results} shows the measured characteristics for 1T, 2T, and 3T models.
We compare the models using the Expected Log Predictive Density (ELPD) which indicate the 3-temperature model is significantly preferred over 2-temperature or 1-temperature models with p(3T)=0.9975.
Discussion of the ELPD can be found in \citet{Vehtari2015}. 

Taking the results of the 3-temperature model, we measure an unspotted photosphere temperature, $\rm T_{phot}$, of $3891\pm37$K with filling factor $f_{\rm phot}=0.64\pm0.05$.
We also measure a cool component attributed to starspots, $\rm T_{spot}$, to be $3020\pm69$K with filling factor $f_{\rm spot}=0.33\pm0.05$.
This and the photospheric component are in excellent agreement with the measurements provided in \citet{Waalkes2024}.
A third temperature is detected near 6000 K. We don't expect any contributions at this temperature as it is much warmer than expected facular temperatures and much cooler than estimated flare temperatures.
It's possible this term is compensating for a wavelength dependent systematic towards the blue end where the S22/G102 light curves are noticeably more affected by poor S/N and unresolved systematics.

These best-fit results from this OOT SED analysis are used to fix the spot and photospheric temperatures for the white light curve analysis and inform the gaussian priors for the atmospheric retrievals. Further information on the SED fits can be found in the Appendix.

\begin{table*}[ht]
\centering
    \begin{tabular}{lcc}											
    \toprule											
    Parameter	& Value$_{-(q50-q16)}^{+(q84-q50)}$ (Spot Model) & Value$_{-(q50-q16)}^{+(q84-q50)}$ (Linear Model)\\
    \midrule			
    {\bf Ramp Parameters} & & \\
    r$_1$	&	$	14.28	_{-	3.23	}^{+	2.91	}$  & 	$	12.55	_{-	2.27	}^{+	2.46	}$  \\
    r$_2$	&	$	-6.82	_{-	0.19	}^{+	0.23	}$  & 	$	-6.58	_{-	0.10	}^{+	0.10	}$	\\
    r$_3$	&	$	0.0045	_{-	0.0016	}^{+	0.0013	}$  & 	$	-0.0101	_{-	0.0007	}^{+	0.0006	}$	\\
    {\bf Breathing Parameters} & &  \\
    b$_1$	&	$	-0.003	_{-	0.020	}^{+	0.018	}$  &   $	-0.060	_{-	0.020	}^{+	0.022	}$  \\
    b$_2$	&	$	-0.003	_{-	0.053	}^{+	0.059	}$  &   $	0.187	_{-	0.068	}^{+	0.061	}$  \\
    b$_3$	&	$	-0.001	_{-	0.075	}^{+	0.068	}$  &   $	-0.218	_{-	0.079	}^{+	0.088	}$	\\
    b$_4$	&	$	0.003	_{-	0.032	}^{+	0.035	}$  &   $	0.093	_{-	0.042	}^{+	0.037	}$	\\
    {\bf Other} & &  \\
    $\beta^{\dagger}$	&  $	0.19	_{-	0.03	}^{+	0.04	}$  & $	0.135	_{-	0.045	}^{+	0.049	}$	 \\
    Slope   &   &   $-0.03083_{-	0.00024	}^{+	0.00024	}$	\\
    \midrule
    {\bf Transit Parameters} & & \\
    $\rm a/R_*$	&   $19.15 $    &	$19.15$	\\
    $\rm Rp/R*$	&	$	0.04656	_{-	0.00023	}^{+	0.00021	}$  &	$0.04886	_{-	0.00018	}^{+	0.00019	}$	\\
    t$_0$ (BJD)	&   $	2459455.98973	_{-	0.00007	}^{+	0.00007	}$  &   $2459455.98971	_{-	0.00007	}^{+	0.00007	}$	\\
    u$_1$	&	$	0.28	_{-	0.04	}^{+	0.03	}$  &	  $0.23	_{-	0.07	}^{+	0.05	}$	\\
    u$_2$	&	$	0.19	_{-	0.05	}^{+	0.06	}$  &	  $0.20	_{-	0.07	}^{+	0.09	}$	\\
    \midrule
    {\bf Stellar Parameters} & & \\
    $\rm T_{Phot}$ (K)  &   3891 & \\
    $\rm T_{Spot}$ (K)  &   3020 & \\
    Limb Spot Latitude  &	$	0.9 $	& \\
    Limb Spot Longitude	&	$	-1.249	_{-	0.034	}^{+	0.058	}$  &	\\
    Limb Spot Radius	&	$	0.276	_{-	0.010	}^{+	0.015	}$  &	\\
    Chord Spot 1 Latitude	&	$	1.469	_{-	0.011	}^{+	0.017	}$  &   \\
    Chord Spot 1 Longitude	&	$	0.065	_{-	0.007	}^{+	0.008	}$  &	\\
    Chord Spot 1 Radius	&	$	0.274	_{-	0.017	}^{+	0.012	}$  &	\\
    Chord Spot 2 Latitude	&	$	1.641	_{-	0.037	}^{+	0.046	}$  &	\\
    Chord Spot 2 Longitude	&	$	0.968	_{-	0.020	}^{+	0.017	}$  &	\\
    Chord Spot 2 Radius	&	$	0.070	_{-	0.037	}^{+	0.040	}$  &	\\
    \bottomrule											
    \end{tabular}																	
\caption{Fixed and measured parameters from white light curve sampling for visit F21/G141. $\dagger$: The $\beta$ parameter is log(error inflation) such that the white light curve uncertainty is multiplied by $10^{\beta}$.}
\label{tab:F21-whitelight}
\end{table*}

\begin{table*}[ht]
    \centering
    \begin{tabular}{lcc}											
    \toprule											
    Parameter	&	Value$_{-(q50-q16)}^{+(q84-q50)}$ (Spotted Model) & Value$_{-(q50-q16)}^{+(q84-q50)}$ (Linear Model) \\
    \midrule
    {\bf Ramp Parameters} & & \\
    r$_1$	&	$	14.149	_{-	2.229	}^{+2.436	}$	&	$	17.50	_{-	4.72	}^{+	4.61	}$	\\
    r$_2$	&	$	-6.580	_{-	0.094	}^{+0.099	}$	&	$	-7.24	_{-	0.17	}^{+	0.16	}$	\\
    r$_3$	&	$	0.0153	_{-	0.0022	}^{+0.0020	}$	&	$	0.0355	_{-	0.0009	}^{+	0.0008	}$	\\
    {\bf Breathing Parameters} & & \\
    b$_1$	&	$	-0.0097	_{-	0.0165	}^{+0.0155	}$	&	$	-0.0427	_{-	0.0172	}^{+	0.0184	}$	\\
    b$_2$	&	$	-0.0334	_{-	0.0482	}^{+0.0516	}$	&	$	0.0353	_{-	0.0571	}^{+	0.0529	}$	\\
    b$_3$	&	$	0.0560	_{-	0.0694	}^{+0.0648	}$	&	$	-0.0569	_{-	0.0705	}^{+	0.0766	}$	\\
    b$_4$	&	$	-0.0300	_{-	0.0318	}^{+0.0340	}$	&	$	0.0307	_{-	0.0374	}^{+	0.0345	}$	\\
    {\bf Other} & & \\
    $\beta^{\dagger}$	&	$	0.29	_{-	0.03	}^{+0.03	}$    &  $	0.25	_{-	0.04	}^{+	0.04	}$	\\
    Slope   &   &   $	-0.0246	_{-	0.0002	}^{+	0.0002	}$	\\
    \midrule
    {\bf Transit Parameters} & & \\
    $\rm a/R_*$	&	$	19.15 $&	$	19.15	$	\\
    $\rm Rp/R*$	&	$	0.0488	_{-	0.0003	}^{+0.0003	}$	&	$	0.04974	_{-	0.00031	}^{+	0.00031	}$	\\
    t$_0$ (BJD)	&	$	2459684.49646_{-0.00009	}^{+0.00009	}$	&	$2459684.49624_{-	0.00007}^{+	0.00007}$	\\
    u$_1$	&	$	0.347_{-0.083}^{+0.065}$	&	$	0.243	_{-	0.115	}^{+	0.094	}$	\\
    u$_2$	&	$	0.229_{-0.081}^{+0.111}$	&	$	0.214	_{-	0.137	}^{+	0.168	}$	\\
    \midrule
    {\bf Stellar Parameters} & & \\
    $\rm T_{Phot}$ (K)  &   3891 & \\
    $\rm T_{Spot}$ (K)  &   3020 & \\
    Limb Spot Latitude  &	$	0.9 $	& \\
    Limb Spot Longitude	&	$	-1.303	_{-	0.002	}^{+0.002	}$	&  \\
    Limb Spot Radius	&	$	0.278	_{-	0.004	}^{+0.004	}$	&	\\
    Chord Spot 1 Latitude	&	$	1.595	_{-	0.022	}^{+0.017	}$	&	\\
    Chord Spot 1 Longitude	&	$	0.056	_{-	0.010	}^{+0.013	}$	&	\\
    Chord Spot 1 Radius	&	$	0.126	_{-	0.016	}^{+0.021	}$	&	\\
    Chord Spot 2 Latitude	&	$	1.699	_{-	0.018	}^{+0.019	}$	&	\\
    Chord Spot 2 Longitude	&	$	0.850	_{-	0.006	}^{+0.006	}$	&	\\
    Chord Spot 2 Radius	&	$	0.059	_{-	0.011	}^{+0.013	}$	&	\\
    \bottomrule											
    \end{tabular}																	
    \caption{Fixed and measured parameters from white light curve sampling for visit S22/G102. }
    \label{tab:S22-whitelight}
\end{table*}

\begin{table*}[htb]
    \centering

\begin{tabular}{lccccc}																									
\toprule																													
$\lambda_{\rm c} (\mu\rm m)$	&	$\rm R_p/R_*$			&	u$_1$			&	u$_2$			&	Slope			&	Error Factor	\\
\midrule
0.80317	&	0.05173	$\pm$	0.00141	&	0.33	$\pm$	0.08	&	0.31	$\pm$	0.09	&	-0.0304	$\pm$	0.0011	&	1.0	\\
0.81107	&	0.04970	$\pm$	0.00137	&	0.36	$\pm$	0.07	&	0.32	$\pm$	0.09	&	-0.0334	$\pm$	0.0010	&	1.0	\\
0.81898	&	0.04911	$\pm$	0.00137	&	0.34	$\pm$	0.08	&	0.30	$\pm$	0.09	&	-0.0350	$\pm$	0.0010	&	1.0	\\
0.85060	&	0.05091	$\pm$	0.00050	&	0.38	$\pm$	0.06	&	0.27	$\pm$	0.08	&	-0.0378	$\pm$	0.0004	&	1.3	\\
0.88485	&	0.05173	$\pm$	0.00074	&	0.25	$\pm$	0.08	&	0.33	$\pm$	0.09	&	-0.0356	$\pm$	0.0006	&	1.0	\\
0.89473	&	0.05051	$\pm$	0.00098	&	0.18	$\pm$	0.06	&	0.23	$\pm$	0.09	&	-0.0299	$\pm$	0.0007	&	1.0	\\
0.90791	&	0.05032	$\pm$	0.00066	&	0.29	$\pm$	0.06	&	0.37	$\pm$	0.08	&	-0.0294	$\pm$	0.0005	&	1.3	\\
0.92168	&	0.05019	$\pm$	0.00080	&	0.26	$\pm$	0.08	&	0.26	$\pm$	0.10	&	-0.0307	$\pm$	0.0006	&	1.0	\\
0.93307	&	0.05004	$\pm$	0.00061	&	0.26	$\pm$	0.07	&	0.33	$\pm$	0.09	&	-0.0289	$\pm$	0.0005	&	1.0	\\
0.94436	&	0.05012	$\pm$	0.00079	&	0.31	$\pm$	0.07	&	0.30	$\pm$	0.09	&	-0.0286	$\pm$	0.0006	&	1.0	\\
0.95955	&	0.05035	$\pm$	0.00049	&	0.29	$\pm$	0.07	&	0.24	$\pm$	0.10	&	-0.0264	$\pm$	0.0004	&	1.1	\\
0.97474	&	0.04961	$\pm$	0.00084	&	0.23	$\pm$	0.08	&	0.34	$\pm$	0.09	&	-0.0239	$\pm$	0.0006	&	1.2	\\
0.98233	&	0.04953	$\pm$	0.00076	&	0.28	$\pm$	0.08	&	0.27	$\pm$	0.10	&	-0.0263	$\pm$	0.0006	&	1.0	\\
0.99941	&	0.05029	$\pm$	0.00046	&	0.30	$\pm$	0.07	&	0.31	$\pm$	0.09	&	-0.0279	$\pm$	0.0003	&	1.2	\\
1.01713	&	0.04896	$\pm$	0.00077	&	0.35	$\pm$	0.07	&	0.29	$\pm$	0.09	&	-0.0233	$\pm$	0.0006	&	1.2	\\
1.03042	&	0.04914	$\pm$	0.00050	&	0.30	$\pm$	0.07	&	0.31	$\pm$	0.09	&	-0.0234	$\pm$	0.0004	&	1.0	\\
1.04307	&	0.04894	$\pm$	0.00072	&	0.30	$\pm$	0.08	&	0.28	$\pm$	0.10	&	-0.0249	$\pm$	0.0005	&	1.0	\\
1.05067	&	0.04861	$\pm$	0.00074	&	0.31	$\pm$	0.08	&	0.27	$\pm$	0.10	&	-0.0206	$\pm$	0.0004	&	1.0	\\
1.05826	&	0.04856	$\pm$	0.00082	&	0.33	$\pm$	0.07	&	0.29	$\pm$	0.09	&	-0.0206	$\pm$	0.0004	&	1.2	\\
1.06543	&	0.04907	$\pm$	0.00077	&	0.30	$\pm$	0.08	&	0.29	$\pm$	0.10	&	-0.0238	$\pm$	0.0006	&	1.0	\\
1.08568	&	0.04920	$\pm$	0.00040	&	0.32	$\pm$	0.06	&	0.32	$\pm$	0.09	&	-0.0251	$\pm$	0.0003	&	1.3	\\
1.10593	&	0.04935	$\pm$	0.00074	&	0.25	$\pm$	0.08	&	0.31	$\pm$	0.10	&	-0.0242	$\pm$	0.0005	&	1.0	\\
1.11268	&	0.05002	$\pm$	0.00076	&	0.31	$\pm$	0.07	&	0.31	$\pm$	0.09	&	-0.0211	$\pm$	0.0005	&	1.0	\\
1.12323	&	0.04933	$\pm$	0.00061	&	0.34	$\pm$	0.07	&	0.27	$\pm$	0.09	&	-0.0202	$\pm$	0.0002	&	1.3	\\
\midrule																			
1.15882	&	0.04879	$\pm$	0.00051	&	0.20	$\pm$	0.06	&	0.20	$\pm$	0.07	&	-0.0313	$\pm$	0.0002	&	1.0	\\
1.18555	&	0.04850	$\pm$	0.00084	&	0.17	$\pm$	0.05	&	0.19	$\pm$	0.07	&	-0.0311	$\pm$	0.0003	&	1.3	\\
1.20265	&	0.04780	$\pm$	0.00068	&	0.16	$\pm$	0.05	&	0.18	$\pm$	0.06	&	-0.0313	$\pm$	0.0003	&	1.0	\\
1.22010	&	0.04864	$\pm$	0.00071	&	0.15	$\pm$	0.04	&	0.18	$\pm$	0.06	&	-0.0304	$\pm$	0.0003	&	1.0	\\
1.23750	&	0.04864	$\pm$	0.00067	&	0.16	$\pm$	0.05	&	0.18	$\pm$	0.06	&	-0.0302	$\pm$	0.0003	&	1.0	\\
1.25545	&	0.04814	$\pm$	0.00067	&	0.14	$\pm$	0.04	&	0.17	$\pm$	0.05	&	-0.0301	$\pm$	0.0003	&	1.0	\\
1.29543	&	0.04854	$\pm$	0.00043	&	0.21	$\pm$	0.07	&	0.26	$\pm$	0.09	&	-0.0286	$\pm$	0.0002	&	1.3	\\
1.33740	&	0.04761	$\pm$	0.00059	&	0.17	$\pm$	0.05	&	0.23	$\pm$	0.07	&	-0.0313	$\pm$	0.0002	&	1.0	\\
1.36983	&	0.04862	$\pm$	0.00043	&	0.27	$\pm$	0.07	&	0.21	$\pm$	0.08	&	-0.0315	$\pm$	0.0002	&	1.0	\\
1.40704	&	0.04891	$\pm$	0.00053	&	0.20	$\pm$	0.06	&	0.19	$\pm$	0.07	&	-0.0336	$\pm$	0.0002	&	1.2	\\
1.43020	&	0.04948	$\pm$	0.00071	&	0.28	$\pm$	0.07	&	0.19	$\pm$	0.08	&	-0.0330	$\pm$	0.0003	&	1.0	\\
1.46032	&	0.04979	$\pm$	0.00040	&	0.32	$\pm$	0.06	&	0.18	$\pm$	0.08	&	-0.0326	$\pm$	0.0002	&	1.0	\\
1.50896	&	0.04984	$\pm$	0.00040	&	0.16	$\pm$	0.05	&	0.25	$\pm$	0.07	&	-0.0321	$\pm$	0.0002	&	1.2	\\
1.54371	&	0.04885	$\pm$	0.00065	&	0.21	$\pm$	0.07	&	0.31	$\pm$	0.09	&	-0.0325	$\pm$	0.0003	&	1.0	\\
1.56073	&	0.04931	$\pm$	0.00067	&	0.23	$\pm$	0.07	&	0.25	$\pm$	0.09	&	-0.0310	$\pm$	0.0003	&	1.0	\\
1.58050	&	0.04760	$\pm$	0.00064	&	0.18	$\pm$	0.06	&	0.25	$\pm$	0.08	&	-0.0301	$\pm$	0.0002	&	1.2	\\
1.60450	&	0.04937	$\pm$	0.00060	&	0.18	$\pm$	0.06	&	0.29	$\pm$	0.09	&	-0.0290	$\pm$	0.0002	&	1.2	\\
1.62850	&	0.05029	$\pm$	0.00056	&	0.23	$\pm$	0.07	&	0.32	$\pm$	0.09	&	-0.0292	$\pm$	0.0002	&	1.0	\\
\bottomrule																													
\end{tabular}

\caption{Transit model results for both visits.}
\label{tab:contaminated-transmission-results}
\end{table*}

\begin{figure*}[ht!]
    \includegraphics[width=0.95\textwidth]{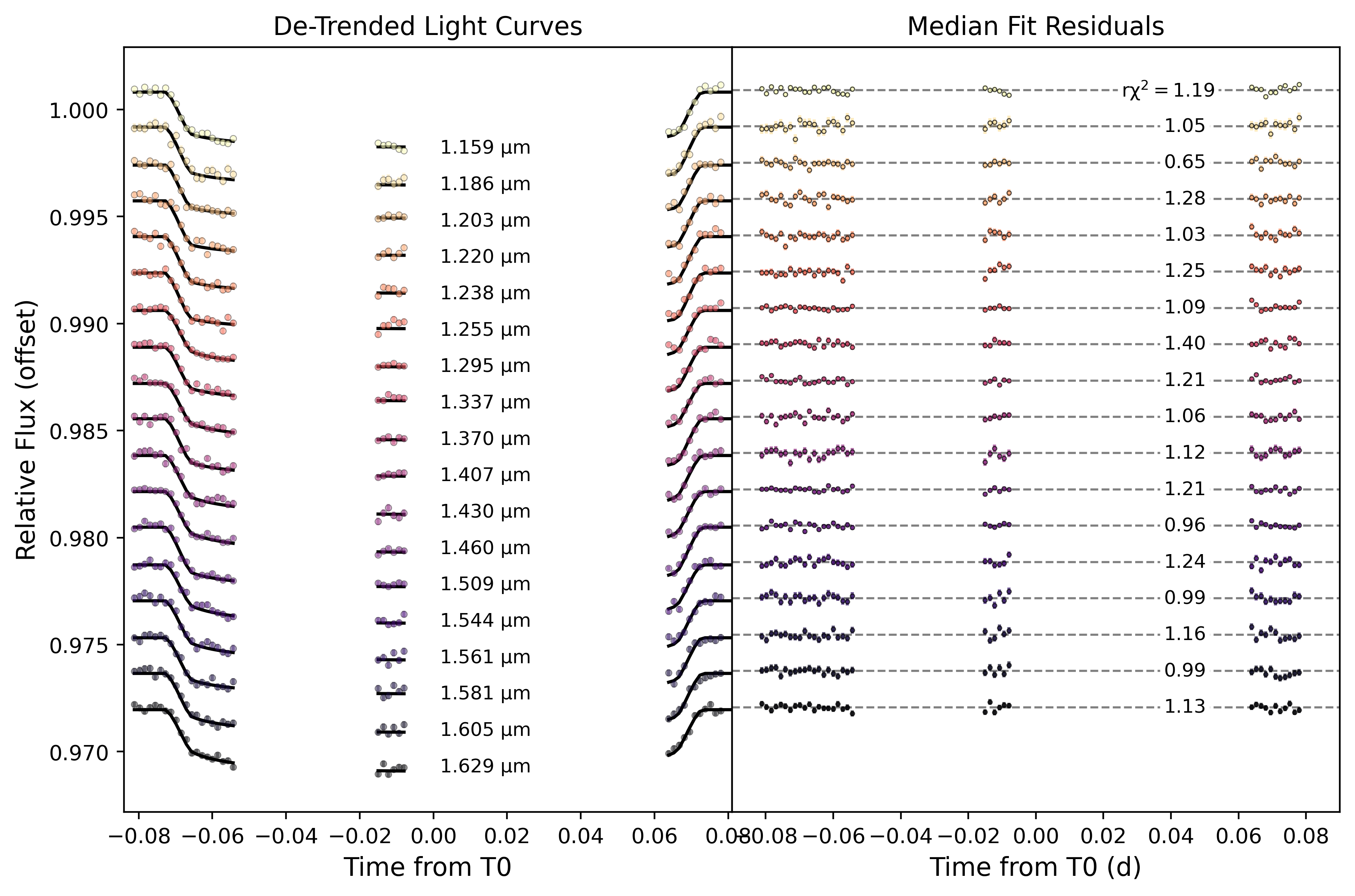}
\caption{F21/G141 light curves with best-fit models, plotted with vertical offsets. $\rm \tilde{\sigma}_{obs}$ represents the median uncertainty for each light curve, with each light curve model's reduced-$\chi^2$ statistic labeled on the right panel. These light curves were systematic-corrected using median parameters from the linear white light curve analysis.}
\label{fig:F21_Lightcurves}
\end{figure*}

\begin{figure*}[ht!]
    \includegraphics[width=0.95\textwidth]{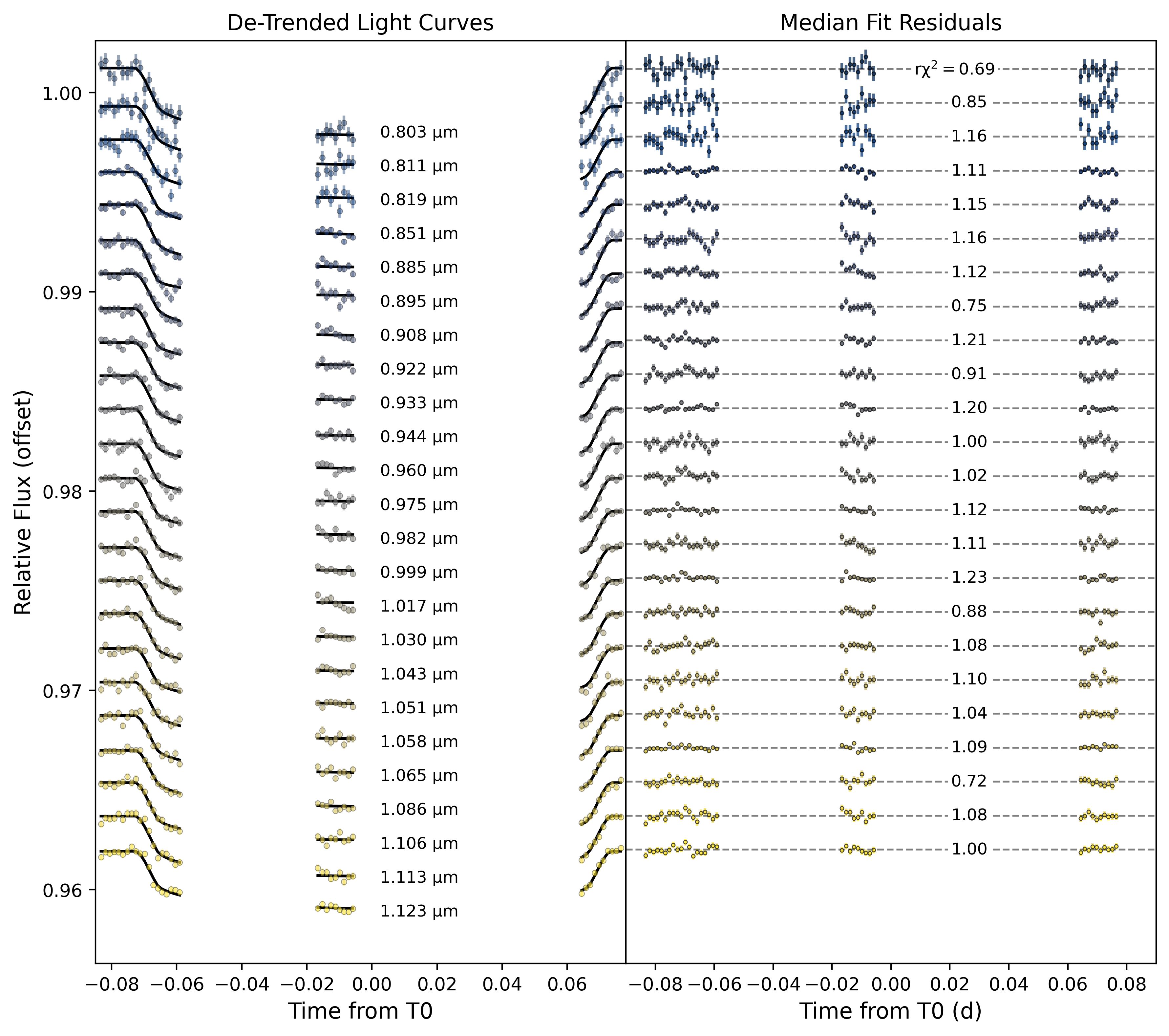}
\caption{Same as Figure \ref{fig:F21_Lightcurves} but for S22/G102.}   
\label{fig:S22_Lightcurves}
\end{figure*}

\subsection{White Light Curves}\label{subsection:results-whitelight}

We analyze both visits separately but find general agreement in the measured parameters between visits, as shown in Figures \ref{fig:star_map}, \ref{fig:white-light-spotted}, and \ref{fig:white-light-linear} as well as Tables \ref{tab:F21-whitelight} and \ref{tab:S22-whitelight}.
Importantly, the spot model results show that the light curves for both visits can be described with approximately the same spot configuration -- an interesting display of long-lived spots considering we are observing the same face of the star in both visits (with 47 stellar rotations between the first and second visit, calculated from the measured mid-transit times in Tables \ref{tab:F21-whitelight} and \ref{tab:S22-whitelight}).
Differences in the measured spot characteristics between visits could be due to spot evolution or the reduced flux contrast in the redder (G141) bandpass.

In both treatments (linear and spotted) of the stellar baseline, there remains additional structure in the light curve residuals that may be due to flares, spots, instrumental effects, or additional forms of stellar activity, and is left unresolved in this analysis.
The instrumental signatures measured by the \textit{linear model} are used for modeling the spectroscopic light curves.

\subsection{Transmission Spectrum}\label{subsection:results-transmission}
We report the measured transit depths of AU Mic b in units of stellar radii in Table \ref{tab:contaminated-transmission-results}.
Binned spectroscopic light curves and residuals calculated from a median correlated set of sampled parameters are shown in Figures \ref{fig:S22_Lightcurves} and \ref{fig:F21_Lightcurves}.
Some apparent orbit-long offsets and correlations are seen in the residuals, which could be due to a number of reasons, including manifestations of the unstable scan that could not be entirely binned out, unresolved stellar activity such as micro-flares, or downstream effects of fixing the breathing parameters to a set of measurements based on the white light curve analysis, which does not ultimately reflect the singularly best-fit set of coefficients.

\subsection{Atmospheric Retrieval}\label{subsection:results-retrieval}

\begin{figure*}[ht!]
\centering
    \includegraphics[width=0.8\textwidth]{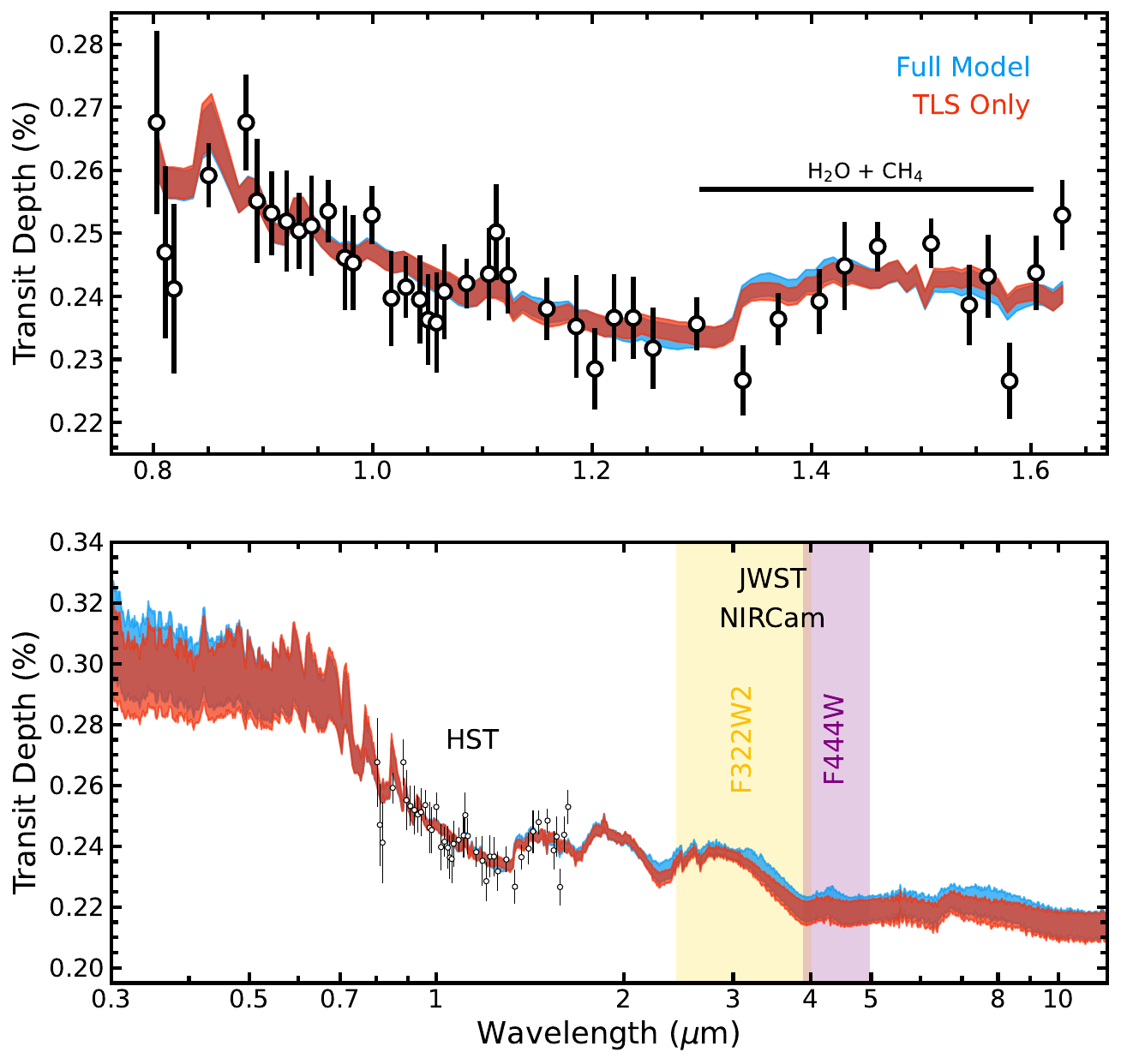}
\caption{Retrieved model spectra in the (top) HST and (bottom) optical and JWST wavelengths for the ``Full Model'' (blue) and ``TLS-Only'' (red) cases. The shaded regions indicate the 1$\sigma$ spread in retrieved spectra. In the top panel the H$_2$O and CH$_4$ absorption bands near 1.4 $\mu$m are indicated, while in the bottom panel the two JWST NIRCam filters (F322W2 in yellow and F444W in purple) that are being used to observe AU Mic b in transmission are indicated.  }   
\label{fig:retrieval}
\end{figure*}

\begin{table*}[tb]
    \centering
    \begin{tabular}{lccc}
        \hline
        \textbf{Parameter Name} & \textbf{Full Model} & \textbf{TLS-Only} & \textbf{Step Function} \\ \hline
        R$_{\rm p}$ (R$_{\oplus}$) @ 1 bar & 3.75$\pm$0.10 & 3.74$^{+0.08}_{-0.07}$ & 4.18$\pm$0.015 \\ 
        T$_{\rm phot}$ (K) & 3891$^{+33}_{-32}$ & 3893$\pm$35 & \nodata \\
        T$_{\rm spot}$ (K) & 3035$^{+60}_{-55}$ & 3038$^{+60}_{-66}$ & \nodata   \\ 
        f$_{\rm spot}$ & 0.36$\pm$0.06 &  0.38$\pm$0.06 & \nodata \\
        $\epsilon'$ (ppm) & 21$^{+15}_{-12}$ & 18$^{+16}_{-12}$ & 49$\pm$14  \\
        $\Delta D$ (ppm) & 30$^{+26}_{-24}$ &  25$^{+22}_{-23}$ & 90$^{+26}_{-25}$ \\
        M$_{\rm p}$ (M$_{\oplus}$) & 36$^{+9}_{-11}$ & \nodata & \nodata  \\
        T$_{\rm atm}$ (K) & 541$^{+251}_{-214}$ & \nodata & \nodata  \\
        ${\rm [M/H]}$ ($\times$ Solar) & 1.5$^{+1.0}_{-1.4}$ & \nodata & \nodata  \\
        C/O & 0.85$^{+0.43}_{-0.45}$ & \nodata & \nodata  \\
        log(P$_{\rm cloud}$) (bar) & -2.4$^{+3.0}_{-3.4}$ & \nodata & \nodata  \\
        log($f_{\rm scatt}$) & 2.0$^{+1.7}_{-1.3}$ & \nodata & \nodata  \\
        $m_{\rm scatt}$ & 9.6$^{+6.6}_{-6.0}$ & \nodata & \nodata  \\
        \hline
        $\chi^2$ & 34.2 & 35.8 & 64.0 \\
        $\chi^2_{\rm r}$ & 1.18 & 1.00 & 1.64 \\
        degrees of freedom & 29 & 36 & 39 \\
        ln(Z) & 336.40 & 336.31 & 327.56  \\
        \hline
    \end{tabular}
    \caption{Median and 1$\sigma$ bounds of the posteriors of our atmospheric retrieval parameters for the ``Full Model'', ``TLS-Only'', and ``Step Function'' cases. The (reduced) chi-squared, degrees of freedom, and log of Bayesian evidence (lnZ) are given at the bottom.}
    \label{tab:platon_results}
\end{table*}

We find the data strongly prefers the ``Full Model'' and ``TLS-Only'' cases over the ``Step Function'' case, with $\Delta$ln(Z)$\sim$8.8 (Table \ref{tab:platon_results}). This is ``decisive'' according to the criteria laid out in \citet{Thorngren2026}, and therefore we can reject a flat spectrum and conclude that we are seeing some signal from the star and/or planet. In contrast, the data does not prefer either of the ``Full Model'' and ``TLS-Only'' cases over the other, as indicated by similarities in their best fit spectra (Figure \ref{fig:retrieval}), ln(Z) and $\chi^2$ values, and retrieved posteriors (Table \ref{tab:platon_results}). In particular, $\Delta$ln(Z)$<$0.1, which is ``barely worth mentioning'' according to \citet{Thorngren2026}. The primary difference between the ``Full Model'' and ``TLS-Only'' cases lie in the G141 wavelengths, where the former exhibits a slightly more pronounced spectral feature than the latter in the H$_2$O and CH$_4$ bands (Figure \ref{fig:retrieval}), leading to a slight decrease in $\chi^2$. However, because there are more varied parameters in the ``Full Model'' case, its $\chi_r^2$ value is actually higher than that of the simpler ``TLS-Only'' case, though both are close to 1. As such, for all intents and purposes, the existing data is insufficient to distinguish between them. 

Most of the parameters that are retrieved for both the ``Full Model'' and ``TLS-Only'' cases are well within 1$\sigma$ of each other, indicating that the underlying atmospheric spectrum does not significantly impact our inferences of the TLS effect from the data. In addition, achieving consistent f$_{\rm spot}$ values between the stellar SED fits and retrievals, as well as reaching $\chi_r^2\sim1$ with tight priors on T$_{\rm phot}$ and T$_{\rm spot}$, which were inferred from the stellar SED, indicates that the stellar SED and the TLS effect in AU Mic b's transmission spectrum are consistent in what they suggest for AU Mic's surface heterogeneity.

\begin{figure*}[ht!]
\centering
    \includegraphics[width=\textwidth]{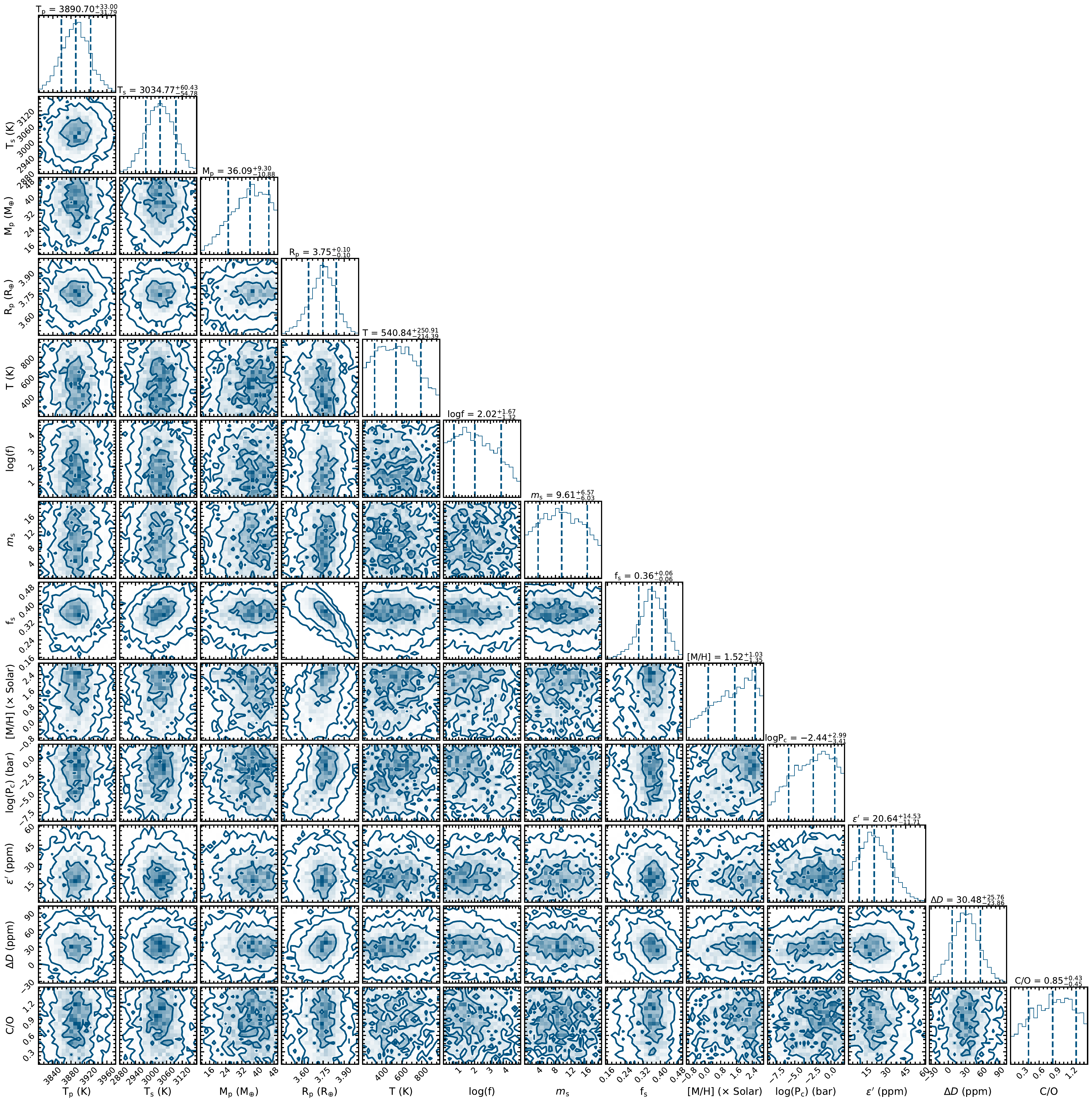}
\caption{Retrieved posteriors for the ``Full Model'' case. Here, T$_{\rm p}$ is T$_{\rm phot}$; T$_{\rm s}$ is T$_{\rm spot}$; T is T$_{\rm atm}$; log(f) is log($f_{\rm scatt}$); $m_{\rm s}$ is $m_{\rm scatt}$; f$_{\rm s}$ is f$_{\rm spot}$; and log(P$_{\rm c}$) is log(P$_{\rm cloud}$).}
\label{fig:platoncorner}
\end{figure*}

The atmospheric parameters inferred in the ``Full Model'' case are generally not well constrained (Figure \ref{fig:platoncorner}). The retrieved mass exceeds the high value presented in \citet{Cale2021} but is likely prior-dependent; when we used a narrower prior as a test ($\mathcal{U}(1,30)$), a lower and more constrained mass posterior resulted, though it remained elevated (22$\pm$6 M$_{\oplus}$). The median isothermal temperature is lower than the expected T$_{\rm eq}$ but is consistent within 1$\sigma$. The median atmospheric metallicity is moderately high ($\sim$30 $\times$ Solar) but can be as low as 1.5 $\times$ Solar within 1$\sigma$. C/O is within 1$\sigma$ of the solar value \citep[0.59;][]{Asplund2021} but can also be $>$1. The cloud top pressure can vary over 6 dex within 1$\sigma$. The haze parameters indicate moderately enhanced scattering compared to Rayleigh scattering, with a scattering slope within 1$\sigma$ of Rayleigh. These results point to a largely unconstrained atmosphere due to the overwhelming TLS effect. 

\section{Discussion}\label{discussion}
In this section we discuss our results, starting with an overview of the caveats of our analysis in \S\ref{subsection:discussion-caveats}. Following those caveats, we briefly discuss what we learn about the stellar photosphere from the white light curve modeling in \S\ref{subsection:discussion-stellarsurface}. Discussion of the atmospheric characterization and implications for this and other young sub-giant planets is in \S\ref{subsection:discussion-atmosphere}.

\subsection{Data Analysis Caveats}\label{subsection:discussion-caveats}
Before discussing and interpreting the results, it is important to recognize the limitations of this analysis: 

\begin{itemize}
    \item We are fundamentally limited by the quality of the observations, which were negatively impacted by an unstable scan across the detector. It is believed the telescope was shaking during these observations, and the resulting time series spectra exhibit a variable PSF that appears as a \textit{lumpiness} across the observations. The irregularity of the effect makes it extremely difficult to mitigate but we attempt to do so by determining an optimal binning scheme, which is ultimately imperfect. 
    \item The stellar spectral models used in this work are based on 1D approximations of quiescent stars, and are poorly descriptive of the 3D convective and radiative flux of spotted stellar surfaces \citep[e.g., ][]{Witzke2022, Smitha2025}. Re-analysis of this dataset, and most like it, will be warranted when improved spectral models (for example, 3D-simulated with MHD) become widely available. 
    \item We use the quadratic limb-darkening law in our transit modeling, which is understood to be a poor description of stellar limb darkening \citep[e.g., ][]{Espinoza2016,Coulombe2024, Keers2024}. Furthermore, it has been demonstrated that this law is particularly poor for M dwarfs in the wavelength regime of HST/WFC3 \citep{Kostogryz2022}. However, these observations are plagued by many issues and a quadratic law appears to be a fine assumption within the precision level of these light curves. Improved limb darkening treatment would be a priority where the instrumental systematics are less severe. 
    \item
    {Flares and Spots on the Light Curve--}
    We trim orbits 3 and 8 from the S22/G102 light curves due to suspected flares, but the exposure time-sampling allowed by HST's orbit leaves gaps throughout the data in which flares may have happened. In that case, the flux of the subsequent orbit may experience an offset \citep[an effect recently described by ][]{Vasilyev2025}, leading to anomalous light curve morphology.
    For these observations, changes in the baseline flux as a result of inter-orbit flares cannot be disentangled from the effects of the variable PSF. 
    
    We chose to model the bump in the egress orbit of S22/G102 as a spot crossing in the white light curve. However, its shape and amplitude does not rule out a flare. Flare morphology is complex and variable \citep{Howard2022ApJ...926..204H,Howard2023ApJ...959...64H,Vasilyev2025, GarciaSoto2025}, and more detailed investigation with higher spectral resolution is likely required to distinguish small spot crossings from flares. 
    We attempted to detect excess flux in hydrogen transition lines around the time of that bump, but the result was inconclusive due to the low spectral resolution. 
\end{itemize}

\subsection{The Spotted Stellar Surface}\label{subsection:discussion-stellarsurface}

After experimenting with different numbers of spots of different size and location on the visible photosphere, we decided the minimum spot configuration justified by the white light curves is three total spots; two spots overlapping the transit chord to fit for occultations and one additional spot on the trailing limb of the star to fit for rotational modulation.
Interestingly, both visits can be modeled with approximately the same three spots, with the most obvious difference being the size of the spot closest to the disk center.
We cautiously suggest that this is a genuinely astrophysical result given that both of these transits occur over nearly the exact same face of the star, with 47 stellar rotations between visits.

Apparent differences in spot size and location between visits may be due to a difference in data quality, the reduced spot contrast in the redder bandpass, or physical evolution of spots on the stellar surface.
It is also important to note that the size and shape of these spots, particularly portions of the spots that are non-overlapping with the transit chord, is unconstrained by this analysis and degenerate with spot contrast.
Because \fleck models spots as circular, we chose to fit the occultations as grazing spots that are not centered on the chord.
Choosing instead to model occultations with more spots of smaller size would produce similar solutions at the expense of computation time.

The inter-visit consistency of the stellar surface as measured only from the white light curves is very promising for characterizing active regions and mapping stellar surfaces using transiting planets, and planned future work which expands this model to the spectroscopic light curves may provide meaningful constraints on the spot spectrum and the total spot coverage of the unocculted disk. 

\subsection{Atmospheric Characterization}\label{subsection:discussion-atmosphere}

\begin{figure}[h]
\centering
    \includegraphics[width=0.45\textwidth]{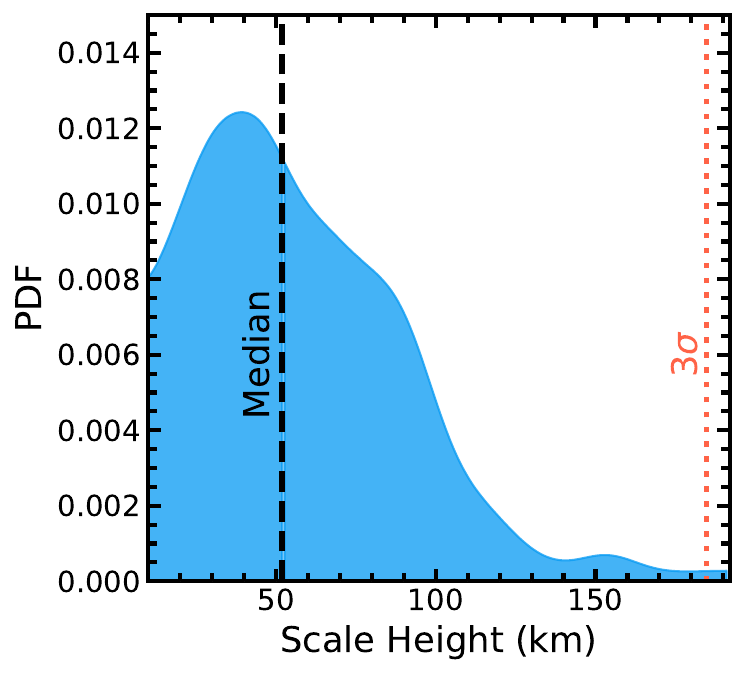}
\caption{Posterior for the scale height of AU Mic b's atmosphere at 1 bar computed from posteriors for M$_{\rm p}$, R$_{\rm p}$, T$_{\rm atm}$, and [M/H] from our ``Full Model'' retrieval. The median (51.8 km) and 3$\sigma$ upper limit (184.8 km) are shown in the black dashed and red dotted lines, respectively.} 
\label{fig:platonsh}
\end{figure}

While our atmospheric inferences are not particularly constraining, the lack of a convincing atmospheric signal puts an upper limit on the ``puffiness'' of AU Mic b. Previously characterized young transiting planets, including those in the V1298 Tau and HIP 67522 systems, exhibit large amplitude spectral features \citep{Barat2024a,Barat2025,Thao2024} that, together with TTV masses \citep{Livingston2026}, point to large atmospheric scale heights of $\sim$1000 km. For comparison, we compute the atmospheric scale height distribution for AU Mic b from our retrieved posteriors for M$_{\rm p}$, R$_{\rm p}$, T$_{\rm atm}$, and [M/H], the latter converted to mean molecular weights using the C/O posterior and the thermochemical equilibrium computations in \texttt{PLATON}. This scale height ``posterior'' is shown in Figure \ref{fig:platonsh} and reveals that, even in the presence of gray clouds and scattering hazes, the data prefer a scale height with a 3$\sigma$ upper limit of only $\sim$185 km. Larger scale heights would result in high amplitude atmospheric features from e.g. H$_2$O and CH$_4$ that would appear in our transmission spectrum and overwhelm the TLS effect.

Our findings suggest that AU Mic b is distinct from the low density planets in the V1298 Tau and HIP 67522 systems, despite their similar ages. While these latter worlds possess relatively low masses for their radii \citep{Thao2024,Livingston2026} and low metallicity atmospheres \citep{Thao2024,Barat2025}, resulting in large scale heights, AU Mic b appears to possess a high mass and/or a high metallicity atmosphere that leads to a scale height only $\sim$1/5 as large. Both V1298 Tau and HIP 67522 are G dwarfs and their planets are expected to lose a significant fraction of their atmospheres and thermally contract in the next $\sim$Gyr, eventually evolving into sub-Neptunes. The potentially much higher mass/metallicity of AU Mic b suggests that it either formed with a much lower gas-to-solid ratio compared to the HIP 67522 and V1298 Tau planets and/or it has lost most of its initially accreted gas mass already. This would fit the emerging picture that low mass planets around M dwarfs may form and evolve differently than those around Sun-like stars \citep{Luque2022,Ho2024}. Alternatively, AU Mic b may host a cloud and haze layer at a much lower pressure level than our retrievals suggest, which would generate a muted atmospheric transmission spectrum even if the scale height were larger than what we have shown in Figure \ref{fig:platonsh}. However, this would still differentiate AU Mic b from the V1298 Tau and HIP 67522 planets that have had atmospheric observations, which showed largely clear atmospheres \citep{Thao2024,Barat2024b}. 

AU Mic b has been observed in transmission by JWST using its NIRCam F322W2 and F444W filters \citep[JWST GO-5311;][]{Feinstein2024jwstprop}, which covers 2.5-5 $\mu$m. The observations have not yet been published, but we can predict AU Mic b's transmission spectrum in the relevant wavelengths by extending our best fit models towards longer wavelengths, as shown in Figure \ref{fig:retrieval}. We find that the ``Full Model'' and ``TLS-Only'' cases remain similar in the F322W2 and F444W filters, with some minor contributions from CH$_4$ and CO$_2$ in the ``Full Model'' case rising slightly above the TLS feature. We therefore predict that the TLS effect is likely to dominate in the JWST observations much like it does in our HST data, and that atmospheric abundances may be difficult to constrain. We also extend our models to shorter wavelengths and find that the optical transit depth of AU Mic b could be nearly 25\% greater than in our HST data due to the TLS effect.

Given the impact of the TLS effect on AU Mic b's transmission spectrum in the absence of an empirical model of the stellar photosphere, emission spectroscopy may be a useful alternative, as it is much less affected by stellar activity. \citet{collins2026arxiv} recently presented a tentative detection of AU Mic b's secondary eclipse in the Spitzer 4.5 $\mu$m band, though the eclipse depth was deeper than expected, yielding a dayside temperature of $\sim$1000 K that is much higher than the estimated equilibrium temperature of $\sim$600 K. The eclipse duration was also longer than predicted by TTV modeling \citep{Wittrock2022,Wittrock2023}. However, if the detection is confirmed, then the precise timing of secondary eclipse would allow for much more efficient observations by JWST in emission in the near- and mid-IR, which could yield vital constraints on its thermal structure and atmospheric chemistry, ultimately leading to a better understanding of low-mass planet formation and early evolution around M dwarfs. 

\section{Conclusions} \label{conclusions}

In this work, we have presented the HST NIR transmission spectrum of AU Mic b, a $\sim$20 Myr Neptune-size planet orbiting a pre-main sequence M dwarf. The dataset was uniquely challenging due to the presence of both typical and atypical HST systematics and several forms of stellar activity. An unstable scan during both visits introduced a variable PSF which we addressed through carefully placed bin edges. Residual effects of this unstable scan are likely the primary limitation in our ability to measure atmospheric absorption in this HST time-series observation. In addition, we attempted to detect and mitigate the effects of flares on our data, but systematics, spots, and flares can have some similarities. Larger bumps in the transit light curve are more likely to be due to spot occultations, but smaller bumps can be attributed to either spots or flares and may require higher spectral resolution to disentangle. Recognizing and mitigating flares in this analysis is severely limited by the discontinuous HST time sampling. Our conclusions are as follows:

\begin{itemize}

    \item We modeled the OOT SED of the host star ( \S\ref{subsection:spec-decomp} and \S\ref{subsection:results-sed}), measuring spot and photosphere temperatures of 3891$\pm$37~K and 3020$\pm$69~K, respectively. The measured spot coverage fraction is $0.33\pm0.05$.
    We fail to detect a hot component that could be attributed to faculae, justifying the assumption that this star's surface is dominated by cool (rather than hot) spots.
    
    \item We fit for the instrumental ramp and breathing effects using the white light curve, taking advantage of the white light curve's $<$50 ppm precision to model the simplest spot configuration required to explain the observations (\S\ref{subsection:systematics-model}, \S\ref{subsection:results-whitelight}, and \S\ref{subsection:discussion-stellarsurface}; Figures \ref{fig:star_map}, \ref{fig:white-light-spotted}, and \ref{fig:white-light-linear}; Tables \ref{tab:F21-whitelight} and \ref{tab:S22-whitelight}). Our best model involves 3 spots--two on the transit chord, and one on the trailing limb that is the source of the visit-long negative slope. Initial characterization of potential spot crossings using the bandpass-integrated white light curves reveals that both visits can be described by a similar distribution of spots. 

    \item We performed atmospheric retrievals on the measured transmission spectrum of AU Mic b, which contains spot contamination (\S\ref{sec:methodretrieval}, \S\ref{subsection:results-retrieval}, and \S\ref{subsection:discussion-atmosphere}; Figures \ref{fig:F21_Lightcurves}, \ref{fig:S22_Lightcurves}, \ref{fig:retrieval}, \ref{fig:platoncorner}, \ref{fig:platonsh}; Tables \ref{tab:binned-lcs}, \ref{tab:platon_priors}, \ref{tab:contaminated-transmission-results}, and \ref{tab:platon_results}). We tested three models: an atmosphere+TLS model, a TLS-only model, and a step function model. We find that the data strongly prefers the atmosphere+TLS and TLS-only models over the step function model, indicating some signal from the atmosphere and/or star. However, the data do not prefer either of the atmosphere+TLS and TLS-only models over the other, resulting in weak atmospheric constraints. In other words, the transmission spectrum can be satisfactorily explained purely by the TLS effect.
    
    \item We used the OOT SED-derived photosphere and spot temperature posteriors as priors in our atmospheric retrievals, but left the spot coverage fraction free to float between 0 and 1. Despite this, our retrievals found spot coverage fractions of 0.36$\pm$0.06 (atmosphere+TLS model) and 0.38$\pm$0.06 (TLS-only model), consistent with that derived from the OOT SED. As such, both the OOT SED and the transmission spectrum offer consistent constraints on the stellar heterogeneity.
    
    \item The retrievals constrain the atmospheric scale height of AU Mic b to be $<$185 km to 3$\sigma$, about 1/5th of those of young planets orbiting G dwarfs like HIP 67522 b and V1298 Tau b. This suggests that AU Mic b possesses a much higher mass given its radii, a much higher atmospheric metallicity, and/or a much greater abundance of optically thick clouds and hazes.
    
    \item Based on our best retrieved models, we predict optical transit depths $\sim$25\% greater than those in our HST data due to the TLS effect. Extending the models towards longer wavelengths, we show that the recent JWST NIRCam F322W2 and F444W observations are likely also dominated by the TLS effect, with potentially weak atmospheric spectral features due to CH$_4$ and CO$_2$. 
\end{itemize}

Our work showcases the challenges facing the characterization of low-mass planets around young M dwarfs and motivates a better treatment of the TLS effect to derived constraints on their atmospheres.
To this end, in Paper II, we will present an empirical model of the stellar photosphere and a physically-motivated method of spot-treating contaminated transmission spectra based on the chromatic characteristics of stellar activity present in the spectroscopic light curves.

\paragraph{Acknowledgments}
Data presented in this publication were accessed through the Mikulski Archive for Space Telescopes (MAST) hosted at the Space Telescope Science Institute.
WCW, PG, and ERN acknowledge support based on observations from HST-GO-15836 provided by NASA through a grant from the Space Telescope Science Institute, which is operated by the Association of Universities for Research in Astronomy, Inc., under NASA contract NAS5-26555.
WCW is supported by the Volkswagen Foundation grant no. 9E126. 
WCW acknowledges support from the TESS Guest Observer Program (80NSSC25K7914).
LA is supported by the Klarman Fellowship.
HRW was funded by UK Research and Innovation (UKRI) framework under the UK government’s Horizon Europe funding guarantee for an ERC Starter Grant [grant number EP/Y006313/1].

\paragraph{Software}
All of the data presented in this article were obtained from the Mikulski Archive for Space Telescopes (MAST) at the Space Telescope Science Institute. The specific observations analyzed can be accessed via \dataset[doi:10.17909/ykar-3e41]{https://doi.org/10.17909/ykar-3e41}.
This project made use of many publicly available tools and packages for which the authors are immensely grateful. In addition to the software cited throughout the paper, we also used \texttt{corner} \citep{corner}, \texttt{JupyterLab} \citep{2013A&A...558A..33A, 2018AJ....156..123A, 2022ApJ...935..167A}, \texttt{Matplotlib} \citep{matplotlib}, \texttt{NumPy} \citep{numpy}, \texttt{Pandas} \citep{pandas}, and \texttt{scipy} \citep{scipy}.

\appendix

\section{SED Fit Results} \label{Appendix-SED}
Here we show the results of the SED spectral decomposition for 2- and 3-temperature models.
The 3-T fit is preferred over 2-T but we lack a robust physical reasoning behind the hot component and it may be correcting for unresolved systematics at the shortest wavelengths. Measured spot and photosphere characteristics are in agreement between the two models.

\begin{figure*}[ht!]
    \centering
    
    \subfloat{\includegraphics[width=0.75\textwidth]{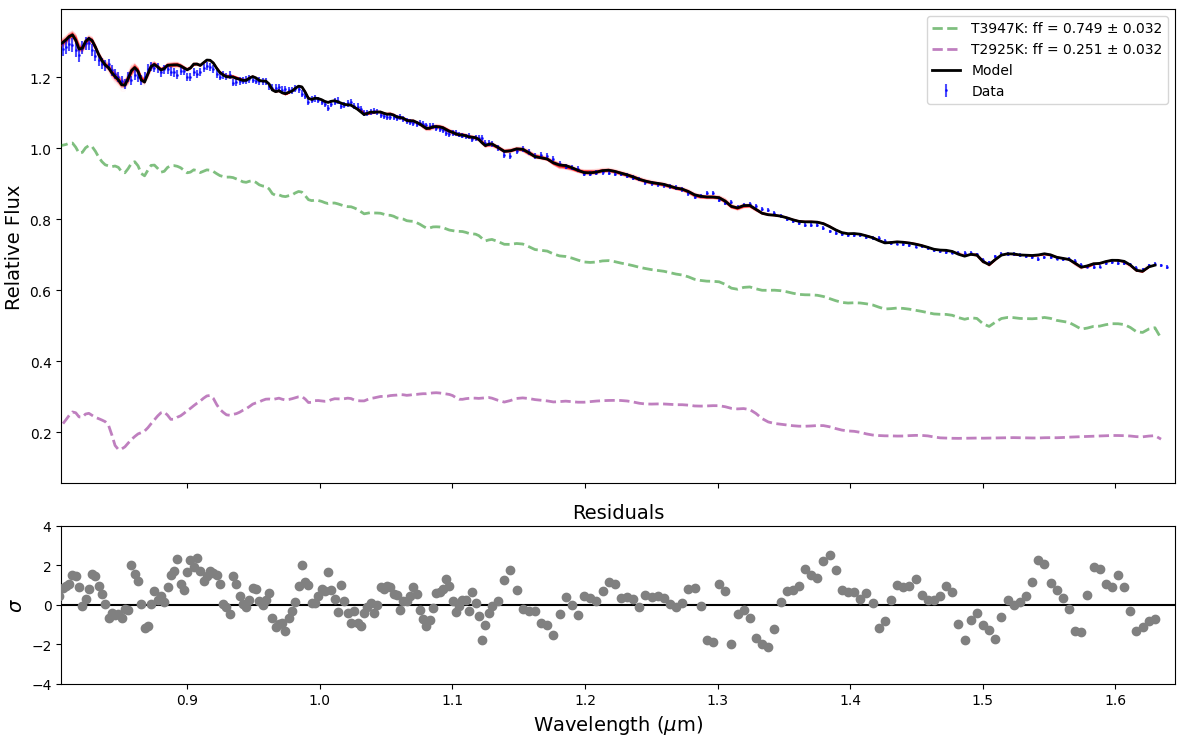}}\\
    \subfloat{\includegraphics[width=0.75\textwidth]{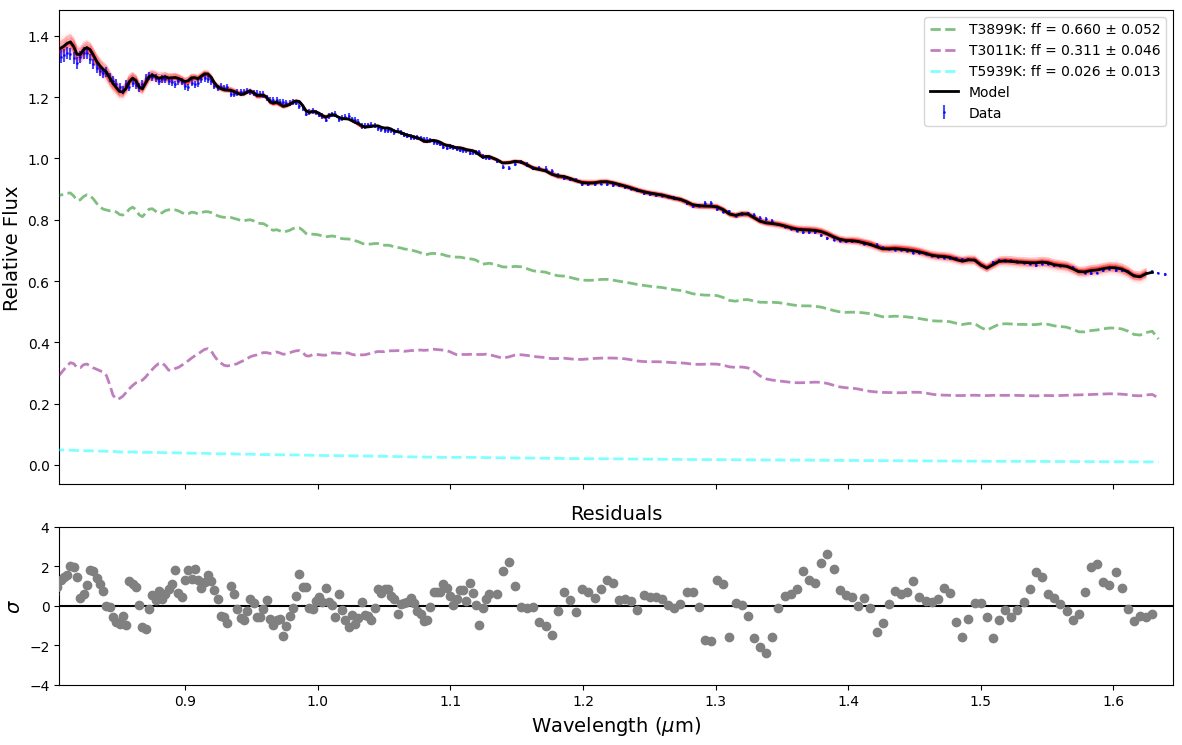}}
    
\caption{SED decomposition with 2 temperature (top) and 3 temperature (bottom) models. Randomly sampled models are plotted in red. }
\label{fig:SED-Models}
\end{figure*}

\begin{figure*}[ht!]
    \subfloat{\includegraphics[width=0.4\textwidth]{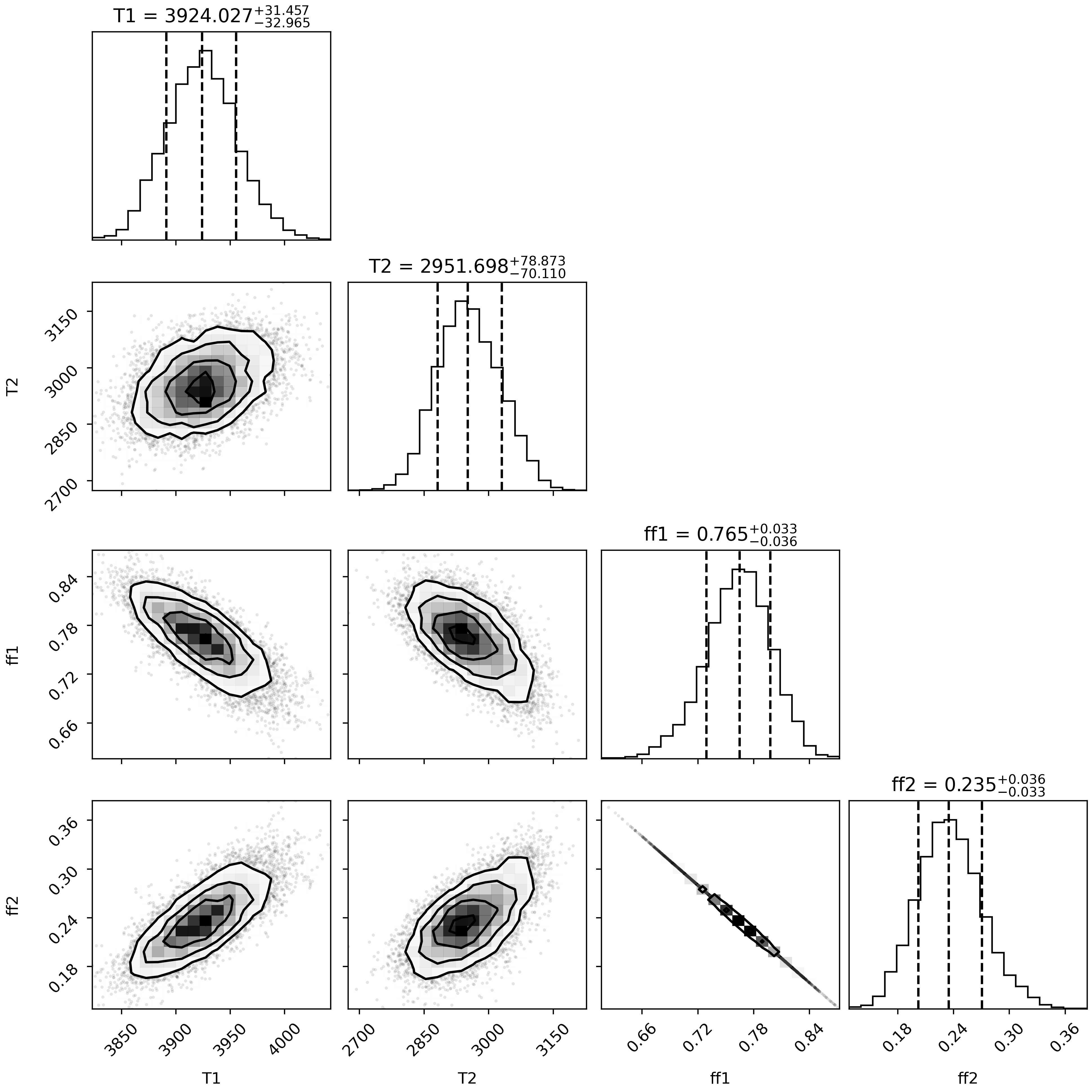}}
    \subfloat{\includegraphics[width=0.6\textwidth]{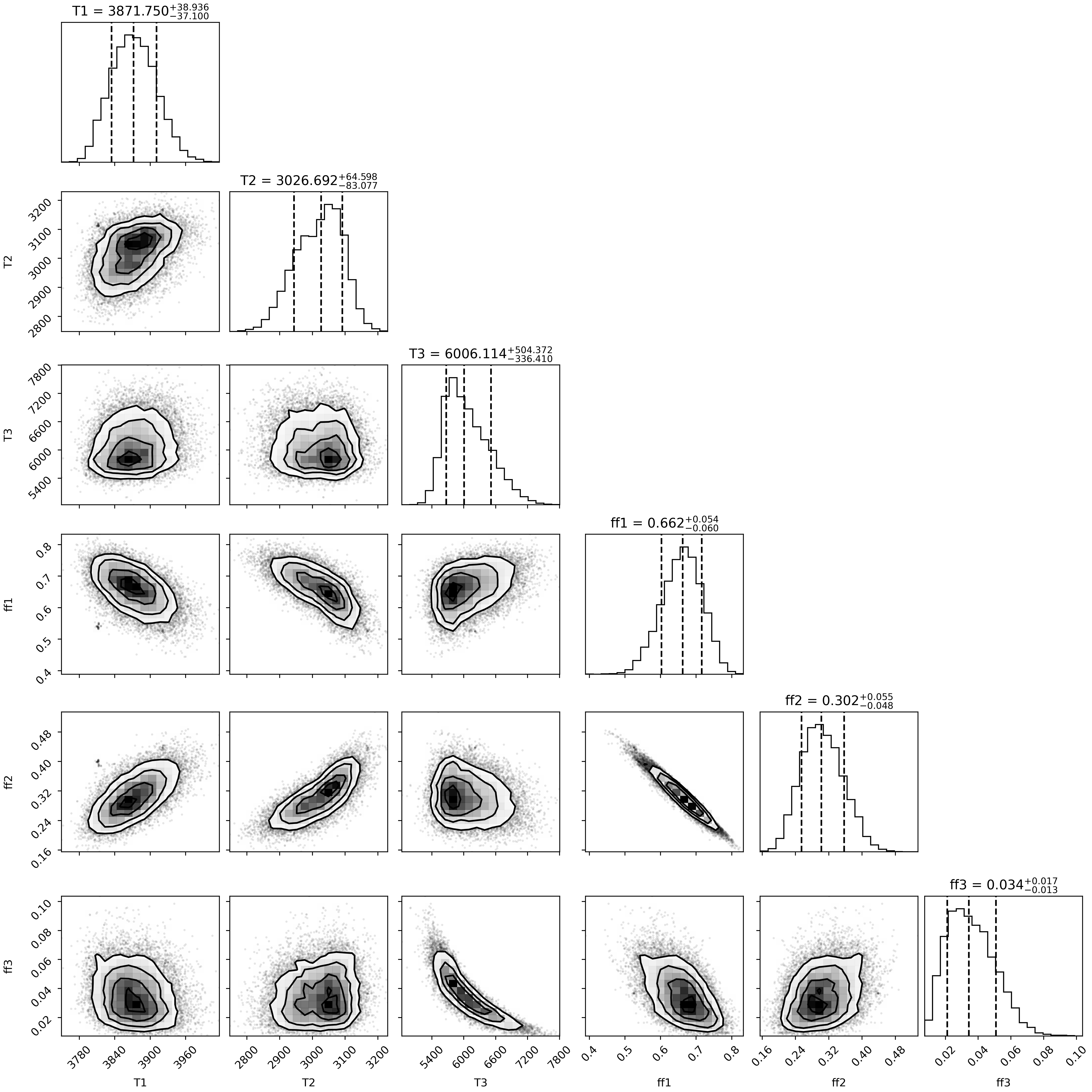}}
\caption{Corner plots for the 2-temperature (left) and 3-temperature (right) SED fits.}
\label{fig:SED-Corner}
\end{figure*}

\bibliography{main.bib}

\end{document}